\documentclass[11pt]{article}

\usepackage{epsfig}

\usepackage{latexsym}
\usepackage{amsfonts}
\usepackage{amsthm}
\usepackage{amsmath}
\usepackage{fullpage}

\newcommand{\vecc}[1]{{\overrightarrow{#1}}}

\newcommand{\old}[1]{{}}
\newcommand{\floor}[1]{{\lfloor #1\rfloor}}
\newcommand{\ceil}[1]{{\lceil #1\rceil}}

\newtheorem{theorem}{Theorem}[section]
\newtheorem{lemma}[theorem]{Lemma}
\newtheorem{conjecture}[theorem]{Conjecture}
\newtheorem{corollary}[theorem]{Corollary}
\newtheorem{claim}[theorem]{Claim}

\newtheorem{fact}[theorem]{Fact}

\theoremstyle{definition}

\begin{document}

\title{On the Reflexivity of Point Sets}
\author{
  Esther M. Arkin%
    \thanks{Department of Applied Mathematics and Statistics,
            Stony Brook University, Stony Brook, NY 11794-3600,
            email: \{\texttt{estie}, \texttt{jsbm}\}\texttt{@ams.sunysb.edu}.}
\and
  S\'andor P. Fekete%
    \thanks{
Department of Mathematical Optimization, 
TU Braunschweig, Pockelsstr.~14,
D-38106 Braunschweig, Germany; \texttt{sandor.fekete@tu-bs.de}.}
\and
  Ferran Hurtado
    \thanks{Departament de Matem\`atica Aplicada II,
            Universitat Polit\`ecnica de Catalunya,
            Pau Gargallo, 5, E-08028, Barcelona, Spain, 
email: \{\texttt{hurtado}, \texttt{noy}, \texttt{vera}\}\texttt{@ma2.upc.es}.}
\and
  Joseph S. B. Mitchell\footnotemark[1]
\and
  Marc Noy\footnotemark[3]
\and
  Vera Sacrist\'an\footnotemark[3]
\and
  Saurabh Sethia
    \thanks{
Department of Computer Science,
Oregon State University, Corvallis, OR 97331, USA. \texttt{saurabh@cs.orst.edu}.
This work was conducted while S.~Sethia was at the Dept. of Computer Science, 
Stony Brook University.}
}

\date{}

\maketitle

\begin{abstract}
  We introduce a new measure for planar point sets $S$ that captures a
  combinatorial distance that $S$ is from being a convex set: The {\em
    reflexivity} $\rho(S)$ of $S$ is given by the smallest number of
  reflex vertices in a simple polygonalization of $S$.  We prove
  various combinatorial bounds and provide efficient algorithms to
  compute reflexivity, both exactly (in special cases) and
  approximately (in general).  Our study considers also some closely
  related quantities, such as the {\em convex cover} number
  $\kappa_c(S)$ of a planar point set, which is the smallest number of
  convex chains that cover $S$, and the {\em convex partition} number
  $\kappa_p(S)$, which is given by the smallest number of convex
  chains with pairwise-disjoint convex hulls that cover $S$.  We have
  proved that it is NP-complete to determine the convex cover or the
  convex partition number and have given logarithmic-approximation
  algorithms for determining each.
\end{abstract}

\section{Introduction}

In this paper, we study a fundamental combinatorial property of a
discrete set, $S$, of points in the plane: What is the minimum number,
$\rho(S)$, of {\em reflex vertices} among all of the {\em simple
  polygonalizations} of $S$?  A {\em polygonalization} of $S$ is a
closed tour on $S$ whose straight-line embedding in the plane defines
a connected cycle without crossings, i.e., a simple polygon.  A vertex
of a simple polygon is {\em reflex} if it has interior angle greater
than $\pi$.  We refer to $\rho(S)$ as the {\em reflexivity} of $S$.
We let $\rho(n)$ denote the maximum possible value of $\rho(S)$
for a set $S$ of $n$ points.

In general, there are many different polygonalizations of a point set
$S$.  There is always at least one: simply connect the points in
angular order about some point interior to the convex hull of $S$
(e.g., the center of mass suffices). A set $S$ has precisely one
polygonalization if and only if it is in convex position; in general,
though, a point set has numerous polygonalizations. Studying the set
of polygonalizations (e.g., counting them, enumerating them, or
generating a random element) is a challenging and active area of
investigation in computational geometry~\cite{aak-eotsp-01,ak-psotd-01,ah-hgrp-96,e-webpage,gnt-lbncf-95,zssm-grpgv-96}.

\begin{figure}
\centerline{\psfig{file=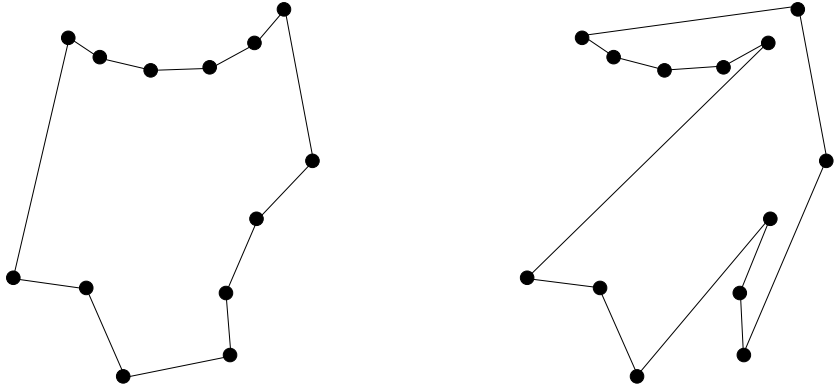,height=2.0in}}
\caption{Two polygonalizations of a point set, one (left) using 7 reflex vertices and one (right) using only 3 reflex vertices.}
\label{fig:example}
\end{figure}

The reflexivity $\rho(S)$ quantifies, in a combinatorial sense, the
degree to which the set of points $S$ is in convex position.  See
Figure~\ref{fig:example} for an example.  We remark that there are
other notions of combinatorial ``distance'' from
convexity of a point set $S$, e.g., the minimum number of points to
delete from $S$ in order that the remaining point set is in convex
position, the number of convex layers, or the minimum number of
changes in the orientation of triples of points of $S$ in order to
transform $S$ into convex position.
  
We have conducted a formal study of reflexivity, both in terms of its
combinatorial properties and in terms of an algorithmic analysis of
the complexity of computing it, exactly or approximately.  Some of our
attention is focussed on the closely related {\em convex cover number}
of $S$, which gives the minimum number of convex chains (subsets of
$S$ in convex position) that are required to cover all points of~$S$.
For this question, we distinguish between two cases: The {\em convex cover
number}, $\kappa_c(S)$, is the smallest number of convex chains required 
to cover $S$; the {\em convex partition number}, 
$\kappa_p(S)$, is the smallest number of convex chains 
with pairwise-disjoint convex hulls required to cover $S$. 
Note that nested chains are feasible for a convex cover
but not for a convex partition.

\paragraph{Motivation.}
In addition to the fundamental nature of the questions and problems we
address, we are also motivated to study reflexivity for several
other reasons:

(1) An application motivating our original investigation is that of
meshes of low stabbing number and their use in performing ray shooting
efficiently.  If a point set $S$ has low reflexivity or a low convex
partition number, then it has a triangulation of low stabbing number,
which may be much lower than the general $O(\sqrt{n})$ upper bound
guaranteed to exist (\cite{a-rsoas-92,hs-parss-95,w-ggssn-93}).
For example, if the reflexivity is $O(1)$, then $S$ has a triangulation
with stabbing number $O(\log n)$.

(2) Classifying point sets by their reflexivity may give us some
structure for dealing with the famously difficult question of counting and
exploring the set of all polygonalizations of $S$. See
\cite{gnt-lbncf-95,zssm-grpgv-96} for some references to this problem.

(3) There are several applications in computational geometry in which
the number of reflex vertices of a polygon can play an important role
in the complexity of algorithms.
If one or more polygons are {\em given} to us, there are many problems
for which more efficient algorithms can be written with complexity in
terms of ``$r$'' (the number of reflex vertices), instead of ``$n$''
(the total number of vertices), taking advantage of the possibility
that we may have $r\ll n$ for some practical instances
(see, e.g., \cite{hm-ftprs-85,hn-tvgrv-96}).
The number of reflex vertices also plays an important role in convex
decomposition problems for polygons; see Keil~\cite{k-pd-00} for a recent
survey, and see Agarwal, Flato, and Halperin~\cite{afg-pdecm-02} for
applications of convex decompositions to computing Minkowski sums of
polygons.

(4) Reflexivity is intimately related to the issue
of convex cover numbers, which
has roots in the classical work of Erd\H{o}s and
Szekeres~\cite{es-cpg-35,es-sepeg-60}, and has been studied
more recently by Urabe~et~al.~\cite{HRU,hu-ndcqp-01,u-opcp-96,u-ppscp-97}.

(5) Our problems are related to some problems in curve (surface)
reconstruction, where the goal is to obtain a ``good''
polygonalization of a set of sample points 
(see, e.g.,~\cite{abe-cbscc-98,dk-spacr-99,dmr-crcdgr-99}).  
  
\paragraph{Related Work.}
The study of convex chains in finite planar point sets is the topic of
classical papers by Erd\H{o}s and
Szekeres~\cite{es-cpg-35,es-sepeg-60}, who showed that any point set
of size $n$ has a convex subset of size $t=\Omega(\log n)$.  This is
closely related to the convex cover number $\kappa_c$, since it
implies an asymptotically tight bound on $\kappa_c(n)$, the worst-case
value for sets of size $n$. There are still a number of open problems
related to the exact relationship between $t$ and $n$; see, for
example, \cite{dcg19} for recent developments.  

Other issues have been considered, such as the existence and
computation (\cite{deo-secp-90}) of large ``empty'' convex subsets
(i.e., with no points of $S$ interior to their hull); this is related
to the convex partition number, $\kappa_p(S)$. It was shown by
Horton~\cite{h-snec7g-83} that there are sets with no empty convex
chain larger than 6; this implies that $\kappa_p(n)\geq n/6$.  

Tighter worst-case bounds on $\kappa_p(n)$ were given by
Urabe~\cite{u-opcp-96,u-ppscp-97}, who shows that $\lceil (n-1)/4
\rceil \leq \kappa_p(n) \leq \lceil 2n/7\rceil$ and that
$\kappa_c(n)=\Theta(n/\log n)$ (with the upper and lower bounds having
a gap of roughly a factor of 2).  (Urabe~\cite{u-ppsdc-99} also studies the
convex partitioning problem in $\Re^3$, where, in particular, the
upper bound on $\kappa_p(n)$ is shown to be $\lceil 2n/9\rceil$.)
Most recently, Hosono and Urabe~\cite{hu-ndcqp-01} have obtained
improved bounds on the size of a partition of a set of points into
disjoint convex quadrilaterals, which has the consequence of improving
the upper bound on $\kappa_p(n)$: $\kappa_p(n)\leq \lceil 5n/18\rceil$
and $\kappa_p(n)\leq (3n+1)/11$ for $n=11\cdot 2^{k-1}-4$ ($k\geq 1$).
The remaining gaps in the constants between upper and lower bounds for
$\kappa_c(n)$ and $\kappa_p(n)$ (as well as the gap that our bounds
exhibit for reflexivity in terms of $n$) all point to the apparently
common difficulty of these combinatorial problems on convexity.

For a given set of points, we are interested in polygonalizations of the
points that are ``as convex as possible''. This has been studied
in the context of TSP (traveling salesperson problem) tours of a
point set $S$, where convexity of $S$ implies (trivially) the
optimality of a convex tour.  Convexity of a tour can be characterized
by two conditions.  If we drop the global condition (i.e., no crossing
edges), but keep the local condition (i.e., no reflex vertices), we
get ``pseudo-convex'' tours.  In \cite{fw-artp-97} it was shown that
any set with $|S|\geq 5$ has such a pseudo-convex tour. It is natural
to require the global condition of simplicity instead, and minimize
the number of local violations -- i.e., the number of reflex vertices.
This kind of problem is similar to that of minimizing the total amount
of turning in a tour, as studied by
Aggarwal~et~al.~\cite{ackms-amtsp-97}.

The number of polygonalizations on $n$ points is, in general,
exponential in~$n$; Garc{\'\i}a et al.~\cite{gnt-lbncf-95} prove a
lower bound of $\Omega(4.64^n)$.

Another related problem is studied by Hosono~et~al.~\cite{HRU}:
Compute a polygonalization $P$ of a point set $S$ such that the
interior of $P$ can be decomposed into a minimum number ($f(S)$) of
empty convex polygons. They prove that $\lfloor (n-1)/4\rfloor \leq
f(n) \leq \lfloor (3n-2)/5\rfloor,$ where $f(n)$ is the maximum
possible value of $f(S)$ for sets $S$ of $n$ points.  The authors
conjecture that $f(n)$ grows like $n/2$.  For reflexivity $\rho(n)$,
we show that $\floor{n/4}\leq \rho(n)\leq \ceil{n/2}$ and conjecture
that $\rho(n)$ grows like $n/4$, which, if true, would imply that
$f(n)$ grows like $n/2$.

We mention one final related problem.  A {\em convex decomposition} of
a point set $S$ is a convex planar polygonal subdivision of the convex
hull of $S$ whose vertices are $S$.  Let $g(S)$ denote the minimum
number of faces in a convex decomposition of $S$, and let $g(n)$
denote the maximum value of $g(S)$ over all $n$-point sets $S$. It has
been conjectured (\cite{RU}) that $g(n)=n+c$ for some constant $c$,
and it is known that $g(n)\le 3n/2$ (\cite{RU}) and that $n+2\le
g(n)$~(\cite{AK}).

\paragraph{Summary of Main Results.}
We have both combinatorial and algorithmic results on
reflexivity.  Our combinatorial results include
\begin{itemize}
\item Tight bounds on the worst-case value of $\rho(S)$ in terms of
  $n_I$, the number of points of $S$ interior to the convex hull of
  $S$; in particular, we show that $\rho(S)\leq \ceil{n_I/2}$ and that
  this upper bound can be achieved by a class of examples.
\item Upper and lower bounds on $\rho(S)$ in terms of $n=|S|$; in
  particular, we show that $\floor{n/4}\leq\rho(n)\leq \ceil{n/2}$.
  In the case in which $S$ has two layers, we show that $\rho(S)\leq
  \ceil{n/4}$, and this bound is tight.
\item Upper and lower bounds on ``Steiner reflexivity'', which is
  defined with respect to the class of polygonalizations that allow
  Steiner vertices (not from the input set $S$).

\end{itemize}
Our algorithmic results include
\begin{itemize}
\item We prove that it is NP-complete to compute the convex cover
  number ($\kappa_c(S)$) or the convex partition number
  ($\kappa_p(S)$), for a given point set~$S$.
\item
  We give polynomial-time approximation algorithms,
having approximation factor $O(\log n)$, for the problems
of computing convex cover number, convex partition number, or
Steiner reflexivity of $S$.
\item
We give efficient exact algorithms to test if $\rho(S)=1$ or $\rho(S)=2$.
\end{itemize}

In Section~\ref{sec:inflection} we study a closely related problem --
that of determining the ``inflectionality'' of $S$, defined to be the
minimum number of inflection edges (joining a convex to a reflex
vertex) in any polygonalization of $S$.  We give an $O(n\log n)$ time
algorithm to determine an inflectionality-minimizing polygonalization,
which we show will never need more than 2 inflection edges.

\section{Preliminaries}

Throughout this paper, $S$ will be a set of $n$ points in the
plane $\Re^2$.  A polygonalization, $P$, of $S$ is a simple polygon
whose vertex set is $S$.  
Let ${\cal P}$ be the set of all polygonalizations of $S$.
Note that ${\cal P}$ is not empty, since any point set $S$ having
$n\geq 3$ points has at least one polygonalization (e.g., the
star-shaped polygonalization obtained by sorting points of $S$
angularly about a point interior to the convex hull of~$S$).

Each vertex of a simple polygon $P$ is either {\em reflex} or {\em
  convex}, according to whether the interior angle at the vertex is
greater than $\pi$ or less than or equal to $\pi$, respectively.  We
let $r(P)$ (resp., $c(P)$) denote the number of reflex (resp., convex)
vertices of $P$.  We define the {\em reflexivity} of a planar point
set $S$ to be $\rho (S) = \min_{P \in {\cal P}} r(P)$.  Similarly, the
{\em convexivity} of a planar point set $S$ is defined to be
$\chi(S)=\max_{P\in{\cal P}} c(P)$.  Note that $\chi(S)=n-\rho(S)$.
We let $\rho(n)=\max_{|S|=n} \rho(S)$.

We let $\textrm{CH}(S)$ denote the {\em convex hull} of $S$.  The point set $S$
is partitioned into (convex) {\em layers}, $S_1,S_2,\ldots$, where the first
layer is given by the set $S_1$ of points of $S$ on the boundary of
$\textrm{CH}(S)$, and the $i$th layer, $S_i$ ($i\geq 2$) is given by the set of
points of $S$ on the boundary of $\textrm{CH}(S\setminus (S_1\cup\cdots\cup
S_{i-1}))$.  We say that $S$ has {\em $k$ layers} or {\em onion depth
  $k$} if $S_k\neq\emptyset$, while $S_{k+1}=\emptyset$.  We say that
$S$ is in {\em convex position} (or forms a {\em convex chain}) if it
has one layer (i.e., $S=S_1$).

A {\em Steiner point} is a point not in the set $S$ that may be added
to $S$ in order to improve some structure of $S$.  We define the {\em
  Steiner reflexivity} $\rho'(S)$ to be the minimum number of reflex
vertices of any simple polygon with vertex set $V\supset S$.  
We let
$\rho'(n)=\max_{|S|=n} \rho'(S)$.
The {\em
  Steiner convexivity}, $\chi'(S)$, is defined similarly.
A {\em convex cover} of $S$ is a set of subsets of $S$ whose union
covers $S$, such that each subset is a convex chain (a set in convex
position).  A {\em convex partition} of $S$ is a partition of $S$ into
subsets each of which is in convex position, such that the convex
hulls of the subsets are pairwise disjoint.  We define the {\em convex
  cover number}, $\kappa_c(S)$,
to be the minimum number of subsets in a convex cover of $S$.
We similarly define the
{\em convex partition number}, $\kappa_p(S)$.
We denote by $\kappa_c(n)$ and $\kappa_p(n)$ the
worst-case values for sets of size~$n$.

Finally, we state a basic property of polygonalizations of point sets.

\begin{lemma} \label{lem:CH-order}
In any polygonalization of $S$,
the points of $S$ that are vertices of the convex hull of $S$ 
are convex vertices of the polygonalization,
and they occur in the polygonalization in the same order
in which they occur along the convex hull.
\end{lemma}

\begin{proof}
  Any polygonalization $P$ of $S$ must lie within
  the convex hull of $S$, since edges of the polygonalization are
  convex combinations of points of $S$.  Thus, if $p\in S$ is a vertex
  of $\textrm{CH}(S)$, then the local neighborhood of $P$ at $p$ lies within a
  convex cone, so $p$ must be a convex vertex of~$P$.

Consider a clockwise traversal of $P$ and let $p$ and $q$ be
two vertices of $\textrm{CH}(S)$ occurring consecutively along $P$.  Then
$p$ and $q$ must also appear consecutively along a clockwise traversal of
the boundary of $\textrm{CH}(S)$, since the subchain of $P$ linking $p$ to $q$
partitions $\textrm{CH}(S)$ into a region to its left (which is outside the
polygon $P$) and a region to its right (which must contain all points of
$S$ not in the subchain).
\end{proof}

\section{Combinatorial Bounds}

In this section we establish several combinatorial results on
reflexivity and convex cover numbers.

\subsection{Reflexivity}

One of our main combinatorial results establishes an upper bound on
the reflexivity of $S$ that is worst-case tight in terms of the number
$n_I$ of points {\em interior} to the convex hull, $\textrm{CH}(S)$,
of $S$.  Since, by Lemma~\ref{lem:CH-order}, the points of $S$ that
are vertices of $\textrm{CH}(S)$ are required to be convex vertices in
any (non-Steiner) polygonalization of $S$, the bound in terms of $n_I$
seems to be quite natural.

\begin{theorem}
\label{thm:reflex-upper-bnd}
Let $S$ be a set of $n$ points in the plane, $n_I$ of which are
interior to the convex hull $\textrm{CH}(S)$.  Then $\rho(S)\leq \ceil{n_I/2}$.
\end{theorem}

\begin{figure}
\centerline{\psfig{file=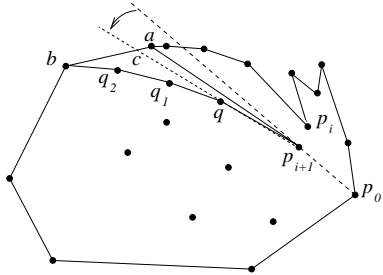,height=1.5in}}
\caption{Computing a polygonalization with at most $\ceil{n_I/2}$ reflex vertices.}
\label{fig:reflex-upper-bnd}
\end{figure}

\begin{proof} We describe a polygonalization in which at most half of
  the interior points are reflex. We begin with the polygonalization
  of the convex hull vertices that is given by the convex polygon
  bounding the hull.  We then iteratively incorporate interior
  points of $S$ into the polygonalization.  Fix a point $p_0$ that
  lies on the convex hull of $S$.  At a generic step of the algorithm,
  the following invariants hold: (1) our polygonalization consists of
  a simple polygon, $P$, whose vertices form a subset of $S$; and (2)
  all points $S'\subset S$ that are not vertices of $P$ lie
  interior to $P$; in fact, the points $S'$ all lie within the
  subpolygon, $Q$, to the left of the diagonal $p_0p_i$, where $p_i$
  is a vertex of $P$ such that the subchain of $\partial P$ from $p_i$
  to $p_0$ (counter-clockwise) together with the diagonal $p_0p_i$
  forms a convex polygon ($Q$).  If $S'$ is empty, then $P$ is a
  polygonalization of $S$ and we are done; thus, assume that $S'\neq
  \emptyset$.  Define $p_{i+1}$ to be the first point of $S'$ that is
  encountered when sweeping the ray $\vecc{p_0p_i}$ counter-clockwise
  about its endpoint $p_0$.  Then we sweep the subray with endpoint
  $p_{i+1}$ further counter-clockwise, about $p_{i+1}$, until we
  encounter another point, $q$, of $S'$.  (If $|S'|=1$, we can readily
  incorporate $p_{i+1}$ into the polygonalization, increasing the
  number of reflex vertices by one.)  Now the ray $\vecc{p_{i+1}q}$
  intersects the boundary of $P$ at some point $c\in ab$ on the
  boundary of~$Q$.
  
  As a next step, we modify $P$ to include interior points $p_{i+1}$ and $q$ (and
  possibly others as well) by replacing the edge $ab$ with the chain
  $(a,p_{i+1}$, $q$, $q_1,\ldots$, $q_k,b)$, where the points $q_i$ are interior
  points that occur along the chain we obtain by ``pulling taut'' the
  chain $(q,c,b)$.  In this ``gift wrapping''
  fashion, we continue to rotate rays counter-clockwise about each interior point
  $q_i$ that is hit until we encounter $b$.  This results in 
  incorporating at least two new interior points (of $S'$) into the
  polygonalization $P$, while creating only one new reflex vertex
  (at~$p_{i+1}$).  It is easy to check that the invariants (1) and (2)
  hold after this step.
\end{proof}

\begin{figure}
\centerline{\psfig{file=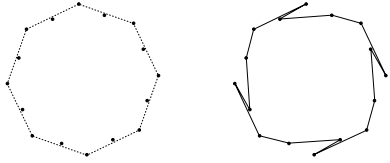,width=2.0in}}
\caption{Left: The configuration of points, $S_0(n)$, which has reflexivity
$\rho(S_0(n))\geq \ceil{n_I/2}$. Right: A polygonalization having $\ceil{n_I/2}$ reflex vertices.}
\label{fig:lower-bnd}
\end{figure}

In fact, the upper bound of Theorem~\ref{thm:reflex-upper-bnd},
$\rho(S)\leq \ceil{n_I/2}$, is {\em tight} in the worst case, as we
now argue based on the special configuration of points, $S=S_0(n)$, in
Figure~\ref{fig:lower-bnd}.  The set $S_0(n)$ is defined for any
integer $n\geq 6$, as follows: $\ceil{n/2}$ points are placed in
convex position (e.g., forming a regular $\ceil{n/2}$-gon), forming
the convex hull $\textrm{CH}(S)$, and the remaining $n_I=\floor{n/2}$
interior points are also placed in convex position, each one placed
``just inside'' $\textrm{CH}(S)$, near the midpoint of an edge of
$\textrm{CH}(S)$.  The resulting configuration $S_0(n)$ has two layers
in its convex hull.

\begin{lemma}
\label{lem:lower-bnd}
For any $n\geq 6$, 
$\rho(S_0(n))\geq \ceil{n_I/2} \geq\floor{n/4}$.
\end{lemma}

\begin{proof}
Let $(x_1,x_2,\ldots,x_{\ceil{n/2}})$ denote the points 
of $S_0(n)$ on the convex hull,  in clockwise order, 
and let $(v_1,v_2,\ldots,v_{\floor{n/2}})$ denote the
remaining points of $S_0(n)$, with 
$v_i$ just inside the convex hull edge $(x_i,x_{i+1})$.
We define $x_{\ceil{n/2}+1}=x_1$.

Consider any polygonalization, $P$, of $S_0(n)$.  {From}
Lemma~\ref{lem:CH-order} we know that the points $x_i$ are convex
vertices of $P$, occurring in the order
$x_1$, $x_2,\ldots$, $x_{\ceil{n/2}}$ around the boundary of $P$.  Consider
the subchain, $\gamma_i$, of $\partial P$ that goes from $x_i$ to
$x_{i+1}$, clockwise around $\partial P$.  Let $m_i$ denote the number
of points $v_j$, interior to the convex hull of $S_0(n)$, that appear
along $\gamma_i$.  

If $m_i=0$, $\gamma_i=x_ix_{i+1}$.  If $m_i=1$, then
$\gamma_i=x_iv_ix_{i+1}$ and $v_i$ is a reflex vertex of $P$; to see
this, note that 
$v_i$ lies interior to the triangle determined by $x_i$, $x_{i+1}$,
and any $v_j$ with $j\neq i$.  If $m_i>1$, then we claim that (a)
$v_i$ must be a vertex of the chain $\gamma_i$, (b) $v_i$ is a convex
vertex of $P$, and (c) any other point $v_j$, $j\neq i$, that is a
vertex of $\gamma_i$ must be a reflex vertex of $P$.  This claim
follows from the fact that the points $x_i$, $x_{i+1}$, and any
nonempty subset of $\{v_j: j\neq i\}$ are in convex position, with the
point $v_i$ interior to the convex hull.  Refer to
Figure~\ref{fig:lower-bnd-proof}, where the subchain $\gamma_i$ is
shown dashed.

\begin{figure}
\centerline{\psfig{file=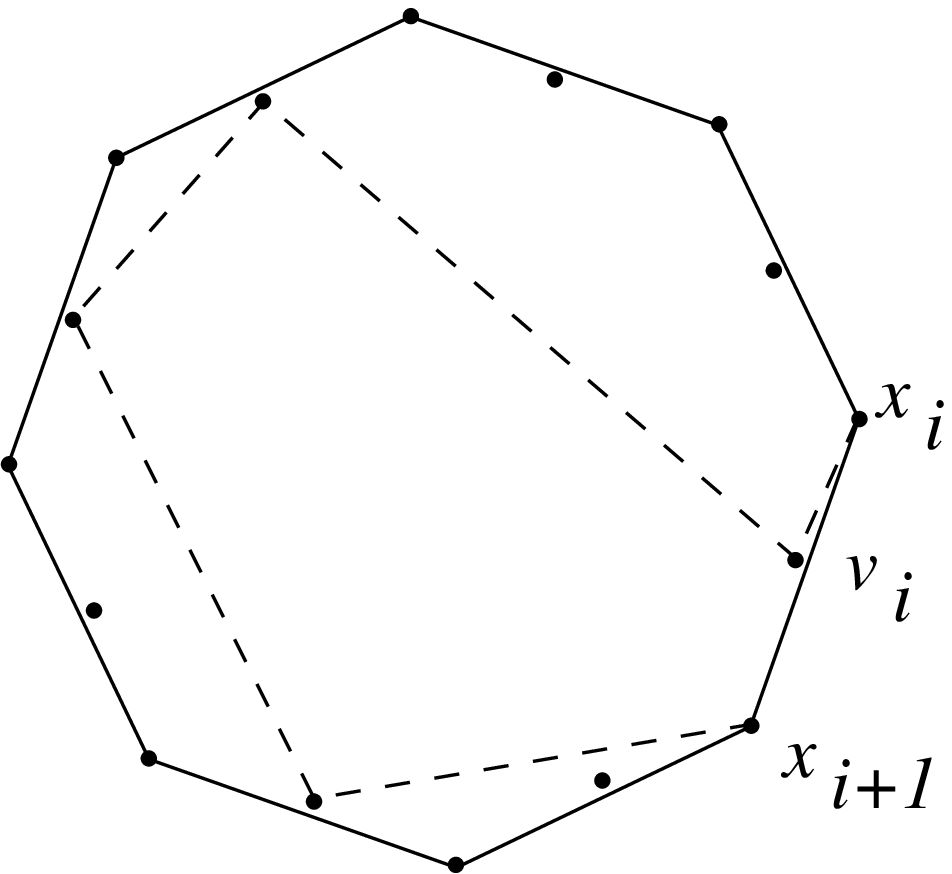,width=1.5in}}
\caption{Proof of the lower bound:
$\rho(S_0(n))\geq \ceil{n_I/2} \geq\floor{n/4}$.}
\label{fig:lower-bnd-proof}
\end{figure}

Thus, the number of reflex vertices of $P$ occurring along $\gamma_i$
is in any case at least $\ceil{m_i/2}$, and we have
\begin{eqnarray*}
\rho(S_0(n)) &\geq& \sum\ceil{m_i/2}\\
&\geq& \left\lceil\sum (m_i/2)\right\rceil=\ceil{n_I/2}\geq\floor{n/4}.
\end{eqnarray*}

\end{proof}

Since $n_I\leq n$, the corollary below is immediate from
Theorem~\ref{thm:reflex-upper-bnd} and Lemma~\ref{lem:lower-bnd}.  The
gap in the bounds for $\rho(n)$, between $\floor{n/4}$ and
$\ceil{n/2}$, remains an intriguing open problem.  While our
combinatorial bounds are worst-case tight in terms of $n_I$ (the number of points
of $S$ whose convexity/reflexivity is not forced by the convex hull of
$S$), they are not worst-case tight in terms of~$n$.

\begin{corollary}
\label{cor:reflex-upper-bnd}
$\floor{n/4}\leq\rho(n)\leq \ceil{n/2}$.
\end{corollary}

Based on experience with a software tool developed by A.~Dumitrescu
that computes, in exponential time, the reflexivity of user-specified
or randomly generated point sets, as well as the proven behavior of
$\rho(n)$ for small values of $n$ (see Section~\ref{sec:small}),
we make the following conjecture:

\begin{conjecture}
\label{con:rho-upper}
$\rho(n)=\floor{n/4}$.
\end{conjecture}

\subsection{Steiner Points}

\begin{figure}
\centerline{\psfig{file=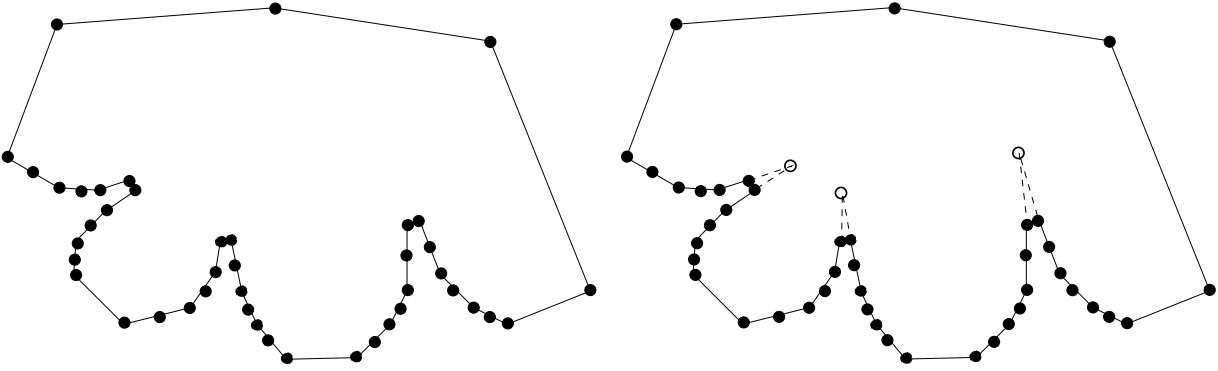,height=1.9in}}
\caption{Left: A point set $S$ having reflexivity $\rho(S)=r$.
Right: The reflexivity of $S$ when Steiner points {\em are} permitted
is substantially reduced from the no-Steiner case: $\rho'(S)=r/2$.}
\label{fig:Steiner-ex}
\end{figure}

If we allow Steiner points in the polygonalizations of $S$, the
reflexivity of $S$ may go down substantially, as the example in
Figure~\ref{fig:Steiner-ex} shows.  In fact, the illustrated class of
examples shows that the use of Steiner points may allow the
reflexivity to go down by a factor of two. 
The Steiner reflexivity, $\rho'(S)$, of $S$ is the minimum number of reflex
vertices of any simple polygon with vertex set $V\supset S$. We conjecture that
$\rho'(S)\geq\rho(S)/2$ for any set $S$, which would imply that this
class of examples (essentially) maximizes the ratio
$\rho(S)/\rho'(S)$.

\begin{conjecture}
\label{con:rho.rho'}
For any set $S$ of points in the plane, $\rho'(S)\geq\rho(S)/2$. 
\end{conjecture}

We have seen (Corollary~\ref{cor:reflex-upper-bnd}) 
that $\floor{n/4}\leq\rho(n)\leq \ceil{n/2}$.  We now show that allowing
Steiner points in the polygonalization allows us to prove
a smaller upper bound, while still being able
to prove roughly the same lower bound:

\begin{theorem}
\label{thm:Steiner-upper}
$$\left\lceil\frac{n-1}{4}\right\rceil-1 \leq \rho'(n)\leq \left\lceil\frac{n}{3}\right\rceil.$$
\end{theorem}

\begin{figure}
\centerline {\input{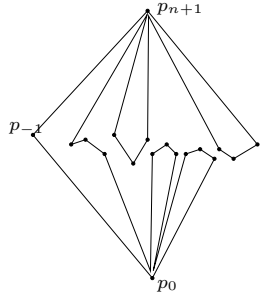}}
\caption{Polygonalization of $n$ points using only $\ceil{n/3}$ 
reflex (Steiner) points. }
\label{fig:Steiner-upper}
\end{figure}

\begin{proof}
  For the upper bound, we give a specific method of constructing a
  polygonalization (with Steiner points) of a set $S$ of $n$ points.
  Sort the points $S$ by their $x$-coordinates and group them into
  consecutive triples. Let $p_{n+1}$ denote a (Steiner) point with a
  very large positive $y$-coordinate and let $p_0$ denote a (Steiner)
  point with a very negative $y$-coordinate.  Each triple, together
  with either point $p_{n+1}$ or point $p_0$, forms a convex
  quadrilateral. Then, we can polygonalize $S$ using one reflex
  (Steiner) point per triple, as shown in
  Figure~\ref{fig:Steiner-upper}, placed very close to $p_{n+1}$ or
  $p_0$ accordingly.  This polygonalization has at most $\ceil{n/3}$
  reflex points.
  
  For the lower bound, we consider the configuration of $n$ points,
  $S$, used in Urabe~\cite{u-opcp-96} to prove that $\kappa_p(n)\geq
  \ceil{(n-1)/4}$.  For this set $S$ of $n$ points, let $P$ be a
  Steiner polygonalization having $r$ reflex vertices.  Then the
  simple polygon $P$ can be partitioned into $r+1$ (pairwise-disjoint)
  convex pieces; this is a simple observation of
  Chazelle~\cite{c-cgc-79} (see Theorem~2.5.1 of \cite{o-cgc-98}).
  The points $S$ occur as a subset of the vertices of these pieces;
  thus, the partitioning also decomposes $S$ into at most $r+1$
  subsets, each in convex position.  Since $\kappa_p(n)\geq \lceil
  (n-1)/4\rceil$, we get that $r\geq \lceil (n-1)/4\rceil -1$.
\end{proof}

\subsection{Two-Layer Point Sets}

Let $S$ be a point set that has two (convex) layers.
It is clear from our repeated use of the example
in Figure~\ref{fig:lower-bnd} that this
is a natural case that is a likely candidate for worst-case behavior.
With a very careful analysis
of this case, we are able to obtain tight combinatorial bounds on
the worst-case reflexivity in terms of~$n$. 

\begin{theorem}
\label{thm:2-layer-bounds}
Let $S$ be a set of $n$ points having two layers.  Then $\rho(S)\leq
\ceil{n/4}$, and this bound is tight in the worst case.
\end{theorem}

\begin{proof}
Consider a set $S$ of $n$ points with onion depth two.  Let $h$ be the
number of points on the convex hull.  Thus there are $n - h$ points on
the interior onion layer.  Let the points on the convex hull be $a_0,
a_1,\dots, a_{h-1}$ in clockwise order.  Let the points on the
interior onion layer be $b_0, b_1,\dots, b_{n - h - 1}$ in clockwise
order.  All arithmetic involving the subscripts of the $a$'s and $b$'s
is done mod $h$ and mod $(n - h)$, respectively.

\begin{fact}
\label{fact:0}
In any polygonalization of $S$ each pocket has at least one
        reflex vertex.
\end{fact}

\begin{proof} 
  This follows from the fact that the polygon defined by the pocket
  and its convex hull edge must have at least three convex vertices
  (as does any simple polygon).
\end{proof}

Consider any point $b_i$ on the interior onion layer.  Let the
intersection of ray $\vecc{b_{i - 1} b_i}$ with the convex hull be $x_i\in
a_j a_{j+1}$.  We call the directed segment $b_i x_i$ the {\em spoke
  $s_i$ originating at $b_i$}, and we say that the spoke $s_i$ belongs
to the convex hull segment $a_j a_{j+1}$ and the segment $a_j a_{j+1}$
has the spoke~$s_i$.  Refer to Figure~\ref{fig:twolayer}.

The number of spokes a convex hull segment has is called its {\em
  spoke count}.  Let $c_j$ be the spoke count of convex hull segment
$a_j a_{j+1}$.  Clearly there are $n - h$ spokes $s_0, s_1,\dots, s_{n
  - h - 1}$ and each spoke belongs to exactly one convex hull segment
(assuming no degeneracy).  Also for all $j$, $0 \leq c_j \leq n-h$ and
$\sum_{j=0}^{h-1} c_j = n - h$.

\begin{figure}[htbp]
\centerline{\input{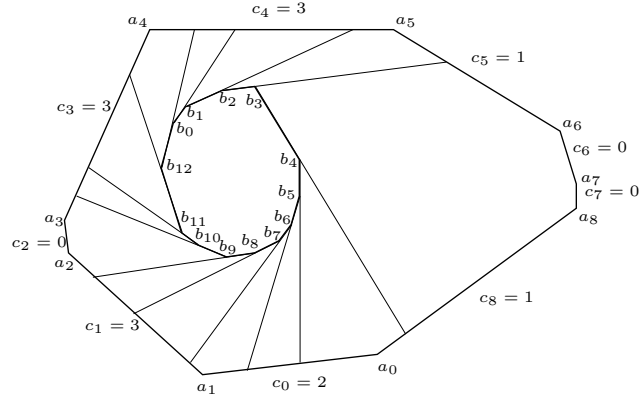}}
\caption{The situation for a point set with two layers.}
\label{fig:twolayer}
\end{figure}

Consider any convex hull edge $a_j a_{j+1}$ with non-zero spoke count
$c_j$.  Assume that the originating points of its spokes are $b_i$,
$b_{i+1}$,$\dots$, $b_{i + c_j - 1}$ in clockwise order.  Then the pocket
$a_j$, $b_i$, $b_{i+1}$,$\dots$, $b_{i + c_j - 1}$, $a_{j+1}$ is called the {\em
  standard pocket} for the convex hull edge $a_j a_{j+1}$.  See
Figure~\ref{fig:standard-pockets}.

\begin{figure}[htbp]
\centerline{\input{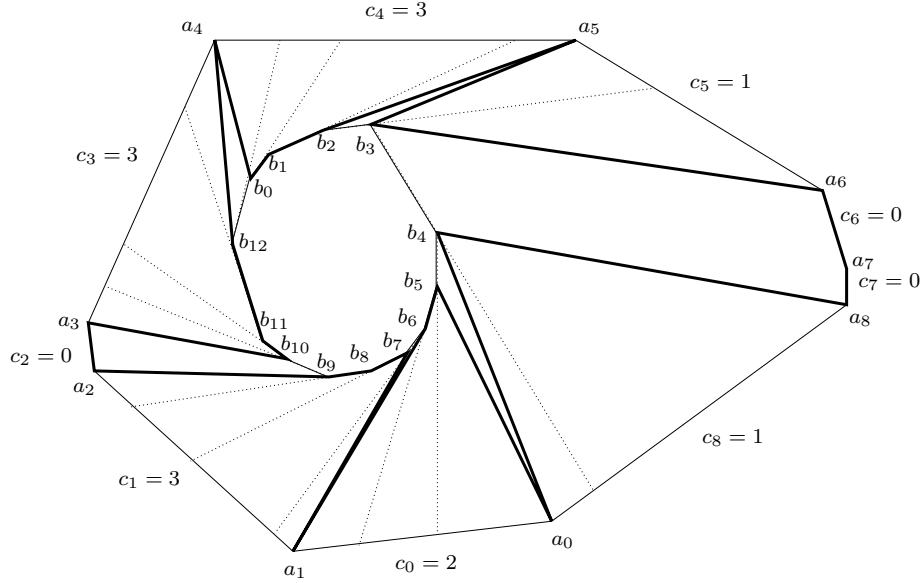}}
\caption{Standard pockets.}
\label{fig:standard-pockets}
\end{figure}

\begin{fact}
\label{fact:1} 
A standard pocket has exactly one reflex vertex.
\end{fact}

\begin{proof}
  Vertices $b_{i+1}, b_{i+2},\dots, b_{i + c_j - 2}$ are convex, since
  the angles at these vertices are the interior angles of the inner
  onion layer.  The vertex $b_{i + c_j - 1}$ is convex because point
  $a_{j+1}$ is on the right of directed line $b_{i + c_j - 2} b_{i +
    c_j - 1}$.  The vertex $b_i$ is reflex because there has to be at
  least one reflex vertex in any pocket.
\end{proof}

\begin{fact}
\label{fact:2}
No two standard pockets intersect each other, except possibly at their
endpoints.
\end{fact}

\begin{proof}
  The annulus between the two onion rings is divided into $n-h$
  disjoint regions by the spokes.  A standard pocket for segment $a_j
  a_{j+1}$ may share a region with each of the standard pockets (if
  any) for segments $a_{j-1} a_j$ and $a_j a_{j+1}$.  In such shared
  regions two pockets have a segment each.  The two segments do not
  intersect because the two segments are obtained by rotating two
  spokes about their points of origin until their intersection with
  the convex hull reaches a vertex.  The two segments can hence either
  remain disjoint or share an endpoint.
\end{proof}

We obtain the {\em standard polygonalization} by connecting all
standard pockets, in order, using convex hull segments with zero spoke
count.  See Figure~\ref{fig:standard-pockets}.

\begin{fact}
\label{lem1}
The number of reflex vertices in the standard polygonalization is at
most $\floor{n/2}$.
\end{fact}

\begin{proof} 
  The number of standard pockets can neither exceed the number of
  convex hull segments $h$ nor the number of internal points, $n-h$.
  Hence, this polygonalization has at most $\min\{h, n-h\}$ standard
  pockets and hence reflex vertices.  This can be at most
  $\floor{n/2}$.
\end{proof}

Consider a convex hull segment $a_j a_{j+1}$ that has a 
non-zero spoke count
$c_j$.  Assume that the origins of its spokes are $b_i, b_{i+1},\dots, b_{i +
  c_j - 1}$.  We call the pocket $a_j$, $b_{i-i}$, $b_i$, $b_{i+1},\dots$,
$b_{i + c_j - 1}$, $a_{j+1}$ the {\em premium pocket} for convex hull
segment $a_j a_{j+1}$.  Note that the premium pocket of a segment has
one more point than its standard pocket.  We obtain the {\em premium
  polygonalization} as follows: Start with the convex hull.
Process convex hull segments in clockwise order beginning anywhere.
If the convex hull segment $a_i a_{i+1}$ has spoke count greater than
or equal to two, replace the edge with its standard pocket.  If the
spoke count is zero, do nothing.  If the spoke count is one, move to
the next segment with non-zero spoke count and replace it with its
premium pocket provided that it is not already processed.  If the next
segment with non-zero spoke count was already processed, replace the
segment being processed with its standard pocket, and we are done.
One can verify that this process gives a valid polygonalization.

\begin{fact}
\label{lem2}
The number of reflex vertices in a premium polygonalization is at most
$\floor{n/3}$.
\end{fact}

\begin{proof}
  In the premium polygonalization, each pocket has at least one reflex
  and one convex vertex, with the exception of the last pocket
  created, which may have only a single reflex vertex.  Thus, the
  number of pockets created cannot exceed $\floor{(n-h+1)/2}$.  Also,
  the number of pockets cannot exceed the number of convex hull edges,
  $h$.  Thus, the number of pockets and, therefore, the number of reflex
  vertices cannot exceed $\min\{h, \floor{(n-h+1)/2}\}$, which can be
  at most $\floor{n/3}$.
\end{proof}

We now define the {\em intruding polygonalization}, as follows: Start
with the convex hull.  Process convex hull segments in clockwise order, 
beginning anywhere.  We consider cases, depending on the spoke count, $c_j$, of
the convex hull segment $a_j a_{j+1}$:
\begin{description}
\item[$c_j=0$] Do nothing.

\item[$c_j\geq 3$] Replace $a_j a_{j+1}$ with a standard pocket.
  
\item[$c_j=1$] Replace the next non-zero count unprocessed hull
  segment with its premium pocket.  (If no such non-zero count
  unprocessed segment exists, then replace $a_j a_{j+1}$ by its
  standard pocket and stop.)

\item[$c_j=2$]  We distinguish two subcases:
\begin{description}
\item[Case A] If this is the last segment to be processed or if the
  sequence of spoke counts following this segment begins with either
  0, or a count $\geq 4$, or an odd number of 1's, or an even number
  of 1's followed by a zero, do the following: Replace $a_j a_{j+1}$
  with its standard pocket.
\item[Case B] If there are an even number of 1's followed by a
  non-zero spoke count, or if the sequence begins with 2 or 3, do one
  of the following, depending on the current ``mode''; initially, the
  mode is ``normal.''
  
  {\bf Normal Mode.} Replace $a_{j+1} a_{j+2}$ with its premium
  pocket.  This leaves one of the originating points of $a_j a_{j+1}$;
  call a point $P$ of this type 
  the {\em pending point}.  Go into ``Point
  Pending Mode.''
  
  {\bf Point Pending Mode.} Let the origins of the spokes of $a_j
  a_{j+1}$ be $b_i, \dots, b_{i + c_j - 1}$.  Find the first point of
  intersection of the ray $\vecc{P b_i}$ with the current polygonalization.
  There are only two choices for where the point of intersection can
  lie.
  
  If the point of intersection lies on $a_j a_{j+1}$, treat ray $\vecc{P
  b_i}$ as a spoke of $a_j a_{j+1}$, thus increasing $c_j$ by 1.  Now
  replace $a_j a_{j+1}$ by its standard pocket, and return to the
  ``Normal Mode.''
  
  If the point of intersection lies on $a_j b_{i-1}$, replace $a_j
  b_{i-1}$ with $a_j b_i P b_{i-1}$.  This reduces $c_j$ from 2 to 1.
  Now handle as in Case~1.
\end{description}
\end{description}

\begin{fact}
\label{lem3}
The intruding polygonalization produces a valid (non-crossing) polygonalization of $S$, with at most $\floor{n/4}$ reflex vertices.
\end{fact}

\begin{proof}
It is straightforward to check each case to make certain that the
polygonalization is valid.
We prove the bound on the number of reflex vertices by charging each
reflex vertex created to a set of 4 input points.

In case $c_j=0$, we do not create any reflex points.

In case $c_j\geq 3$, we create a reflex vertex as part of the standard
pocket.  We charge this to the three internal vertices of the pocket
and the source vertex of the segment being processed.

In case $c_j=1$, we create a reflex vertex as part of the premium
pocket of the next segment having non-zero spoke count.  We charge
this to the sources of the segment being processed and the next
segment and to the (at least two) internal vertices of the pocket.

If $c_j=2$ and we are in Case A, we create a reflex vertex as part of
the standard pocket.  We charge this to the source of the segment
being processed and the two internal vertices.  We need to charge it
to one more point.
\begin{itemize}
\item If the next segment has zero spoke count, we charge it to its
  source.
\item If the next segment has spoke count greater than or equal to
  four, we charge it to one of its internal vertices.  (Recall that
  for a segment with spoke count greater than or equal to four, only
  three of its internal vertices will get charged by its own standard
  pocket.)
\item If there are some number of 1's followed by a 0, we charge the
  source of segment having spoke count 0.
\item If there are odd number of 1's followed by a non-zero count,
  then note that the segment following this sequence of 1's will be
  replaced by its premium pocket by our algorithm.  This segment will
  have a spoke count greater than or equal to 2.  We put the extra
  charge needed on one of the internal points of this premium pocket.
  (Recall that this premium pocket will have at least 3 internal
  points and the pocket itself will charge only two of them.)
\end{itemize}

If $c_j=2$ and we are in Case B, we consider each of the two modes separately:
\begin{description}
\item[Normal Mode] If the sequence begins with a 2 or 3, we charge the
  reflex vertex created to the internal points (at least 3) of the
  premium pocket and the source of segment being replaced.
  
  If the sequence has an even number of 1's followed by a non-zero
  spoke count, then we have a case similar to that above, but we have
  only two internal points to charge in the premium pocket.  However,
  since the first segment with spoke count 1 will be replaced by its
  premium pocket, there will be an odd number of 1's remaining.
  Hence, the segment following this sequence of one (which has spoke
  count $\geq 2$) will be replaced with a premium pocket, which has an
  extra internal point to which we can charge.
  
  Note that in Normal Mode, in both of the cases above, 
  the source of the
  segment being processed remains uncharged.  Also, the pending point
  remains to be incorporated into the polygonalization.
\item[Point Pending Mode] Depending on where ray $\vecc{P b_i}$ intersects in
  this case, there are two possibilities:
  
  If the point of intersection lies on $a_j a_{j+1}$, then we are
  creating a pocket with at least 2 internal points.  We charge the
  reflex vertex created to the two internal points and to the source
  of the segment being processed and to the uncharged point in the
  previous Normal Mode.  See Figure~\ref{fig:case2b1}.

\begin{figure}
\centerline{\input{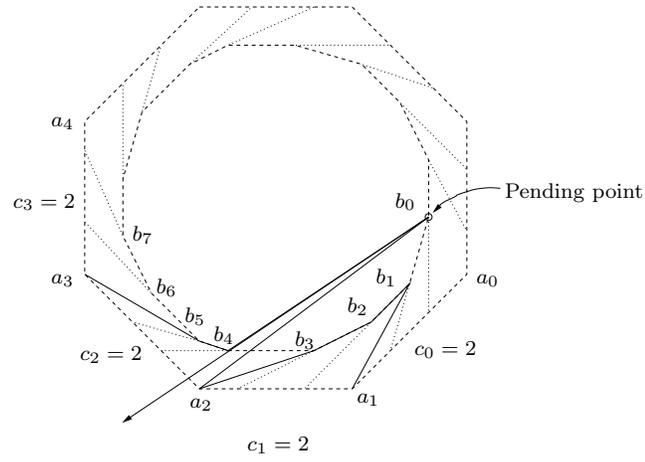}}
\caption{Subcase of Case B, $c_j=2$: Ray ${b_0 b_4}$ intersects segment $a_2 a_3$, but not segment $b_3 a_2$.}
\label{fig:case2b1}
\end{figure}

\begin{figure}
\centerline{\input{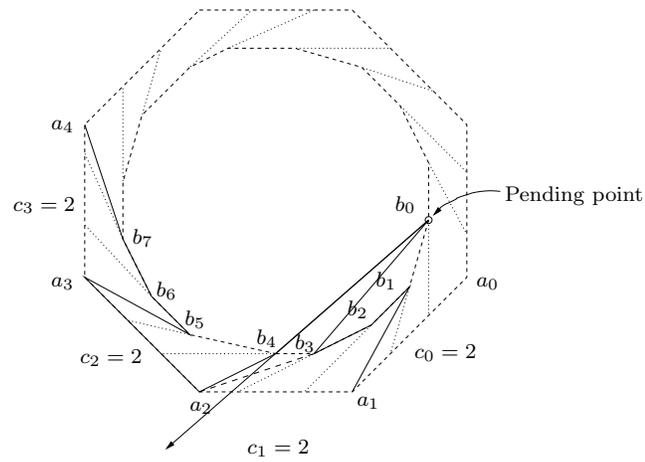}}
\caption{Subcase of Case B, $c_j=2$: Ray ${b_0 b_4}$ intersects segment $b_3 a_2$, but not segment $a_2 a_3$.}
\label{fig:case2b2}
\end{figure}

  If the point of intersection lies on $a_j b_{i-1}$, then the
  subpocket has two internal points.  In this case we
  charge its reflex point
  (which is the pending point $P$) to these two internal points and to
  the uncharged point in the previous normal mode.  We need to make
  one more charge, which depends on the spoke count of next segment.
  If it is 2 or 3, we charge the extra internal point in its premium
  pocket.  If it is 1, note that since there were even number of 1's
  following, after we replace the next one with its premium pocket,
  there would be odd number of 1's remaining.  Hence, the segment
  following this sequence of 1's (which has non-zero spoke count),
  will be replaced with its premium pocket, by our algorithm.  Hence,
  we can charge its extra internal point.  See Figure~\ref{fig:case2b2}.
\end{description}

\end{proof}

This concludes the proof of Theorem~\ref{thm:2-layer-bounds}
\end{proof}

\paragraph{Remark.}
Using a variant of the polygonalization given in the proof of
Theorem~\ref{thm:2-layer-bounds}, it is possible to show that a
two-layer point set $S$ in fact has a polygonalization with at most
$\ceil{n/3}$ reflex vertices such that none of the edges in the
polygonalization pass through the interior of the convex hull of the
second layer.  (The polygonalization giving upper bound of
$\ceil{n/4}$ requires edges that pass through the interior of the
convex hull of the second layer.) This observation may be useful in
attempts to reduce the worst-case upper bound ($\rho(n)\leq
\ceil{n/2}$) for more general point sets~$S$.

\subsection{Convex Cover/Partition Numbers}

As a consequence of the Erd\H{o}s-Szekeres
theorem~\cite{es-cpg-35,es-sepeg-60}, Urabe has given bounds on the
convex cover number of a set of $n$ points:
Urabe~\cite{u-opcp-96} and Hosono and Urabe~\cite{hu-ndcqp-01}
have obtained bounds as well on the convex partition number
of an $n$-point set:
$$\left\lceil\frac{n-1}{4}\right\rceil \leq \kappa_p(n)\leq \left\lceil\frac{5n}{18}\right\rceil.$$

While it is trivially true that $\kappa_c(S)\leq \kappa_p(S)$
the ratio $\kappa_p(S)/\kappa_c(S)$ for a set $S$
may be as large as $\Theta(n)$; the set $S=S_0(n)$ (Figure~\ref{fig:lower-bnd})
has $\kappa_c(S)=2$, but $\kappa_p(S)\geq n/4$.

The fact that $\kappa_p(S)\leq \rho(S)+1$ follows easily by
iteratively adding $\rho(S)$ segments to an optimal polygonalization
$P$, bisecting each reflex angle. The result is a partitioning of $P$
into $\rho(S)+1$ convex pieces.
Thus, we can obtain a convex partitioning of $S$ by associating a
subset of $S$ with each convex piece of $P$, assigning each point of
$S$ to the subset associated with any one of the convex pieces that has
the point on its boundary.  
(This is the same observation of Chazelle~\cite{c-cgc-79} used in
  the proof of Theorem~\ref{thm:Steiner-upper}.)

\begin{figure}[hbtp]
   \begin{center}
   \epsfxsize=.4\textwidth
   \ \epsfbox{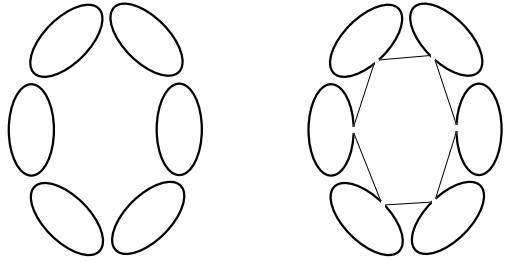}
   \caption{An example with $\rho(S)=2\kappa_p(S)$.
   Each thick oval shape represents a numerous subset of points of $S$
in convex position.}
   \label{fig:ratio-two}
   \end{center}
\end{figure}

We believe that the relationship between reflexivity ($\rho(S)$)
and convex partition number ($\kappa_p(S)$) goes the other way as well:
A small convex partition number should imply a small reflexivity.
In particular, we have invested considerable effort in trying to
prove the following conjecture:
\begin{conjecture}
\label{con:rho.kappa2}
$\rho(S)=O(\kappa_p(S))$. 
\end{conjecture}
The reflexivity can be as large as {\em twice} the
convex cover number ($\rho(S)=2\kappa_p(S)$), as illustrated
in the example of Figure~\ref{fig:ratio-two}; however,
this is the worst class of examples we have found so far.

Turning briefly to Steiner reflexivity, it is not hard to see that
$\rho'(S)=O(\kappa_p(S))$ (see the proof of
Corollary~\ref{cor:reflexivity-approx}). 
Thus, a proof of
Conjecture~\ref{con:rho.kappa2} would follow from the validity of
Conjecture~\ref{con:rho.rho'}.

\subsection{Small Point Sets}
\label{sec:small}

It is natural to consider the exact values of $\rho(n)$,
$\kappa_c(n)$, and $\kappa_p(n)$ for small values of $n$.
Table~\ref{tab:small} below shows some of these values, which we
obtained through (sometimes tedious) case analysis.  Aichholzer and
Krasser~\cite{ak-psotd-01} have recently applied their software that
enumerates point sets of size $n$ of all distinct order types to
verify our results computationally; in addition, they have obtained
the result that $\rho(10)=3$.  (Experiments are currently
under way for $n=11$; values of $n\geq 12$ seem to be
intractable for enumeration.)

\begin{table}
\small
\centerline{%
\begin{tabular}{|l||c|c|c|}
\hline
$n$ & $\rho(n)$ & $\kappa_c(n)$ & $\kappa_p(n)$ \\
\hline \hline
$\leq $ 3 &0 &1 & 1\\
       4 &1 &2 & 2\\
       5 &1 &2 & 2\\
       6 &2 &2 & 2\\
       7 &2 &2 & 2\\
       8 &2 &2 & 3\\
       9 &3 &3 & 3\\
      10 &3 &- &- \\
\hline
\end{tabular}%
}
\vspace{1ex}
\caption{\label{tab:small}
Worst-case values of $\rho$, $\kappa_c$, $\kappa_p$ for small
values of $n$.
}
\end{table}

\section{Complexity}
\label{sec:npc}

We now prove lower bounds on the
complexity of computing the convex cover number, $\kappa_c(S)$, and
the convex partition number, $\kappa_p(S)$.
The proof for the convex cover number uses a reduction of the problem
1-in-3 SAT and is inspired by the hardness proof for the {\sc Angular
  Metric TSP} given in~\cite{ackms-amtsp-97}.  The proof for the
convex partition number uses a reduction from {\sc Planar 3\,Sat}.

\begin{theorem}
\label{th:npc.disjoint}
It is NP-complete to decide whether for a planar point set $S$
the convex partition number $\kappa_p(S)$ is below some
threshold $k$.
\end{theorem}

\begin{proof}
We give a reduction from {\sc Planar 3\,Sat}, which was shown to be
NP-complete by Lichtenstein (see~\cite{l-pftu-82}). A {\sc 3\,Sat} instance $I$
is called a {\sc Planar 3\,Sat} instance, if the (bipartite) 
``occurrence graph''
$G_I=(V_I,E_I)$ is planar, where each vertex of $V_I$ 
corresponds to a variable or a clause, and two vertices
are joined by an edge of $E_I$ if and only if the vertices
correspond to a variable $x$ and a clause $c$ such that $x$ appears 
in the clause $c$ in $I$.
See Figure~\ref{fi:npc.disjoint}(a) for an example, 
where a solid edge denotes
an un-negated literal, while a dashed edge represents a negated literal
in a clause.

\begin{figure}[h!btp]
   \begin{center}
   \epsfxsize=.20\textwidth
\ \epsfbox{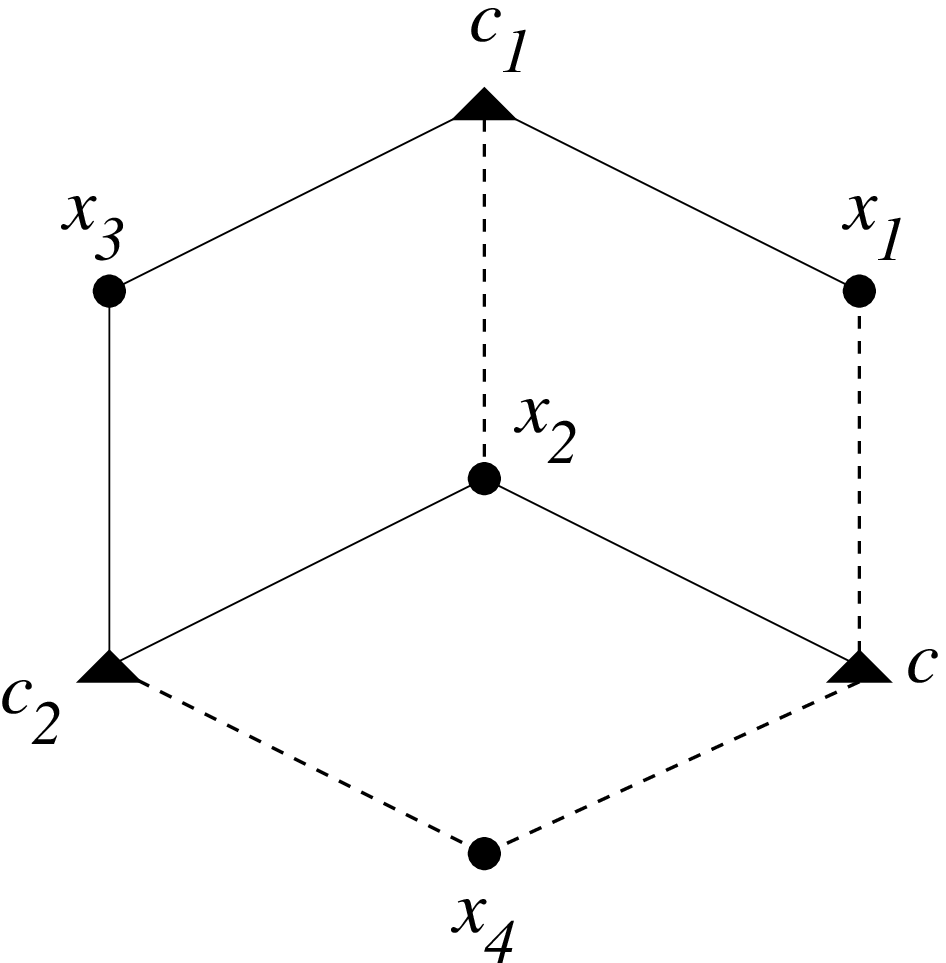}
\hfill
   \epsfxsize=.100\textwidth
\ \epsfbox{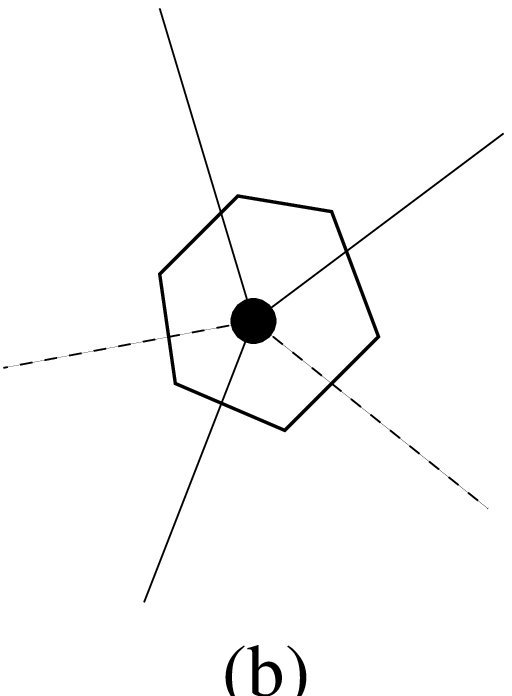}
\hfill
   \epsfxsize=.25\textwidth
\ \epsfbox{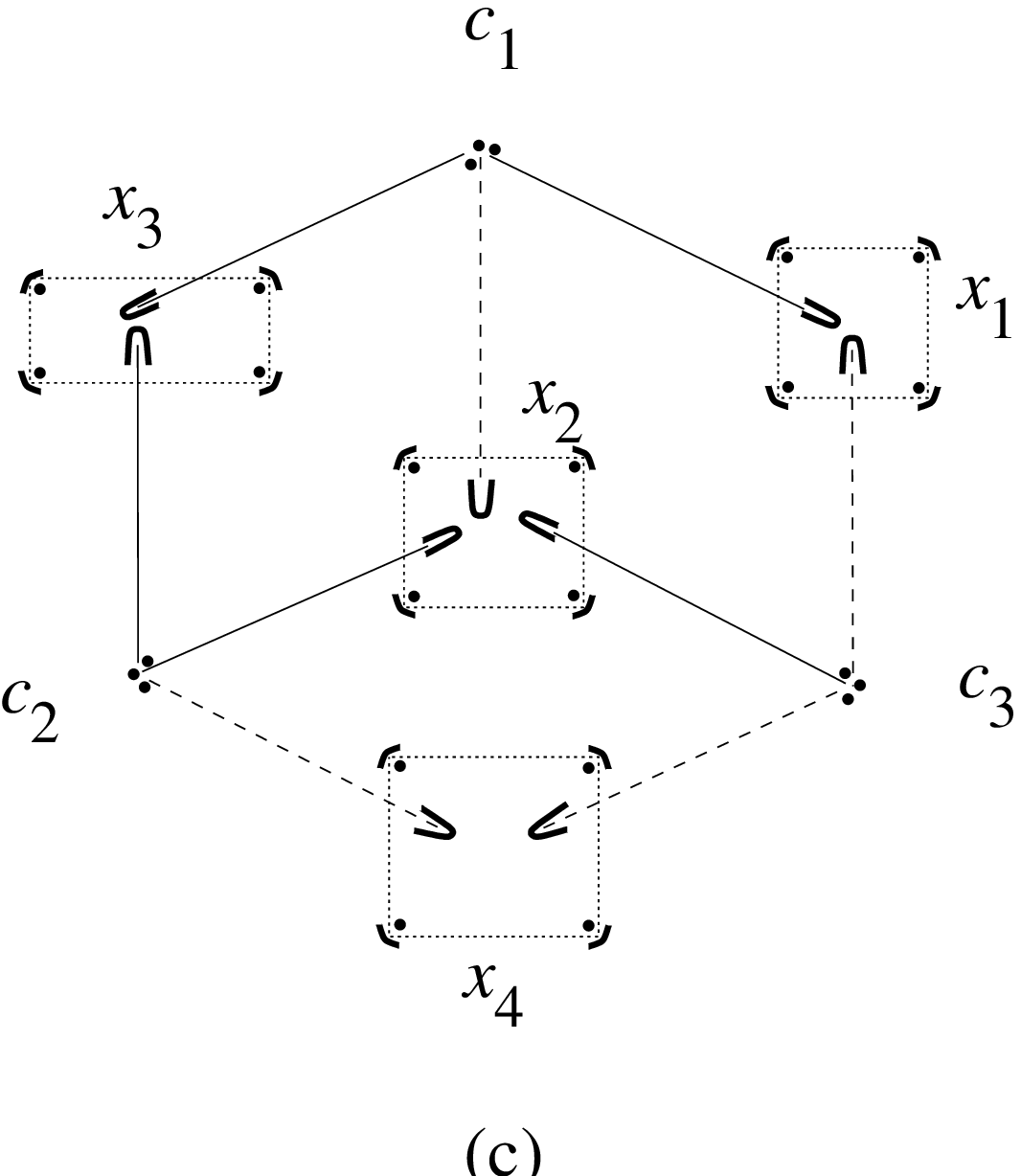}
\hfill
   \epsfxsize=.25\textwidth
\ \epsfbox{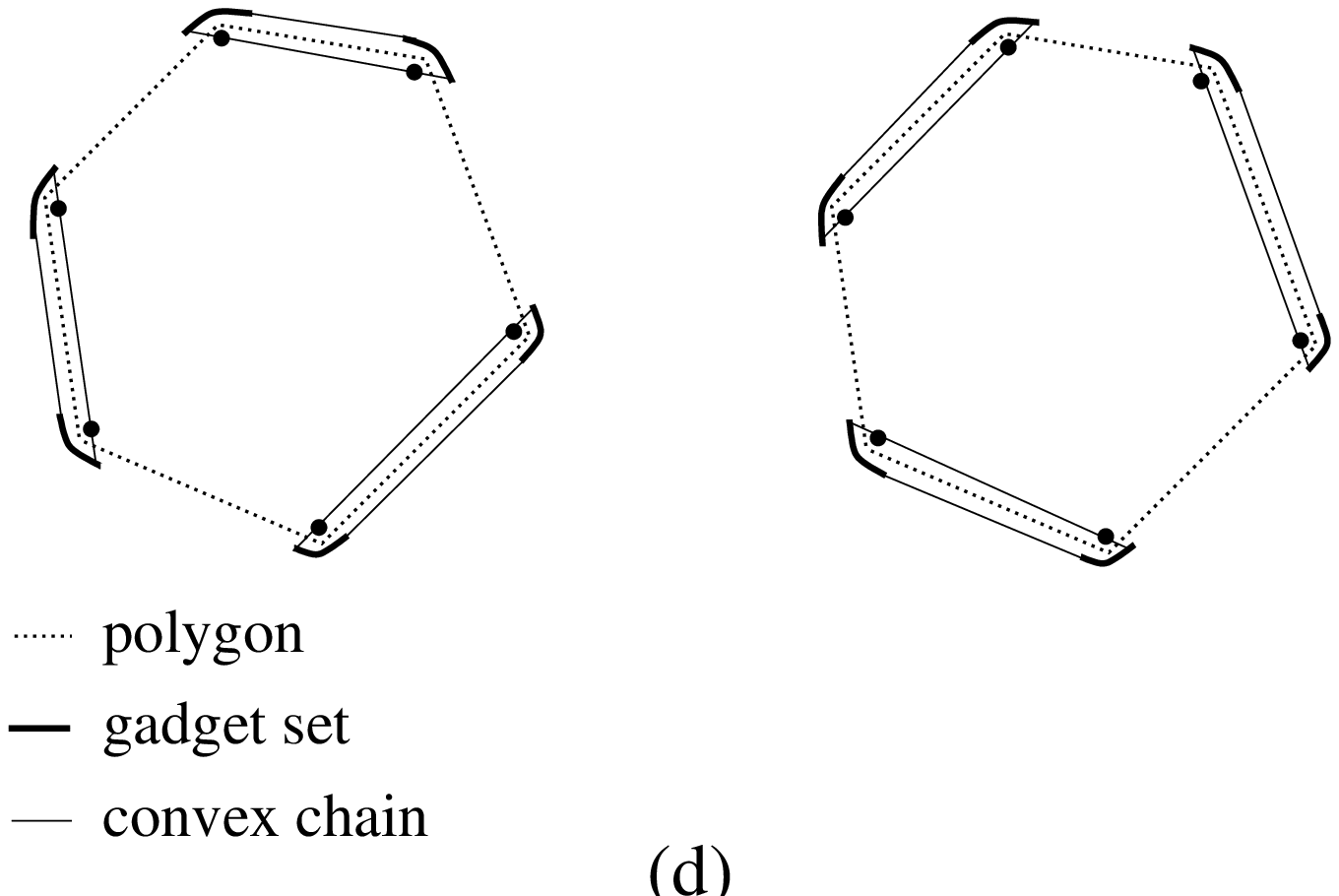}
\caption{(a) A straight-line embedding of the occurrence graph for the 
{\sc 3\,Sat} instance $(x_1\vee\overline{x_2}\vee x_3) \wedge (x_2\vee{x_3}\vee\overline{x_4})\wedge (\overline{x_1}\vee x_2 \vee \overline{x_4})$;
(b) a polygon for a variable vertex;
(c) a point set $S_I$ representing the {\sc Planar 3\,Sat} instance $I$;
(d) joining point sets along the odd or even polygon edges.}
\label{fi:npc.disjoint}
   \end{center}
\end{figure}

The basic idea is the following: Each variable is represented
by a set of points that can be partitioned into $s$ disjoint 
convex chains in two different ways. One of these possibilities 
will correspond to a setting of ``true'', the other to a setting of 
``false''.  Each clause is represented by a set of points, such
that it can be covered by three convex chains disjoint from
all other chains, if and only if at least one of the variables is set
in a way that satisfies the clause.

So let $I$ be a {\sc Planar 3\,Sat} instance with $n$ variables 
and $m=O(n)$ clauses. Consider a straight-line
embedding of its occurrence graph $G_I$. In a first step, for 
every vertex $v_x$ representing a variable $x$, draw 
a small polygon $P_x$ with $2s\leq 2m$ edges 
around $v_x$, where $s$ is bounded by the maximum degree of a variable
vertex. (See Figure~\ref{fi:npc.disjoint}(b).) This is done
such that no edge of $G_I$ passes through one of the corners of $P_x$,
and only edges adjacent to $v_x$ intersect $P_x$; moreover, we
choose edge orientations for all polygons that assure that 
no two different polygons $P_x$ and $P_y$ have two collinear edges
(See Figure~\ref{fi:npc.disjoint}(c), where
we have $s=2$ and we have used rectangles as polygons to keep the figure clear.)
Furthermore, we choose the polygons such that
all of the line segments connecting $v_x$ to clauses where
$v_x$ appears un-negated intersect ``even'' edges of the polygon, 
and all the line segments connecting $v_x$ to clauses
where $v_x$ appears negated intersect ``odd'' edges of the polygon.

In a second step,
replace the polygons (and thus the variable vertices) by
appropriate sets of $2s(kn^4+1)$ points
-- see Figure~\ref{fi:npc.disjoint} for an example.
Each of the corners $b_x^{(1)},\ldots,b_x^{(2s)}$ 
of a polygon $P_x$
is represented by a dense convex chain $B_x^{(i)}$
of $kn^4$ points, forming
a convex curve with an opening of roughly $\pi/s$, and an additional 
``pivot'' point $p_x^{(i)}$. 

Finally, we describe how to represent the clauses:
Each vertex $v_c$ in $G_I$ representing a clause $c$ is replaced by
three points forming a small triangle $T_c$; furthermore, for each
of the three edges connecting some
variable vertex $v_x$ to $v_c$, add a convex chain $C_{x,c}$
of $k*n^2$ points within the polygon $P_x$. $C_{x,c}$
is lined up with the triangle $T_c$
such that $T_c\cup C_{x,c}$ is a convex chain.
On the other hand, the opening of $C_{x,c}$ towards $T_c$
is narrow enough to prevent any other point 
from forming a convex chain with
all points in $C_{x,c}$. We also avoid any other 
collinearities in the overall arrangement. 

\smallskip
Now we claim the following correspondence:

The resulting point set $S_I$ can be partitioned into at
most $3m+sn$ disjoint convex chains, if and only if the {\sc Planar 3\,Sat}
instance $I$ is satisfiable.

\smallskip
To see that a satisfying truth assignment induces a feasible decomposition, 
note that there are exactly two ways to cover the points for a
polygon with $2s$ edges by at most $s$ disjoint convex chains.
(One of these choices arises by joining the pairs of sets that belong
to the ``even'' edges of a polygons box into one convex chain, 
the other by joining the pairs of sets for the ``odd'' edges
of a polygon. This is shown in Figure~\ref{fi:npc.disjoint}(d).)
For a true variable, take the
``even'' choice, for a false variable, the ``odd'' choice.
Now consider a clause $c$ and a variable $x$ that satisfies $c$.
By the choice of chains covering the points for $x$,
we join the triangle $T_c$ for $v_c$ with the chain
$C_{x,c}$ into one convex chain, without intersecting any other chain. 
The other two chains $C_{y,c}$ are each covered by separate chains.
This yields a decomposition into $3m+sn$ disjoint convex chains, as claimed.

\smallskip
To see the converse, assume we have a decomposition
into at most $3m+sn$ disjoint convex chains.

We start by considering how the sets $B_x^{(i)}$ and the sets
$C_{x,c}$
(henceforth called ``gadget sets'' $G_j$, with 
$j=1,\ldots,3m+2sn$) can be covered in such a solution.
Associate each chain with all the $G_j$
of which it covers at least $\Omega(n)$ points.
It is straightforward to see that a convex chain
that covers at least $\Omega(n)$ points from each of three
different gadget sets must contain
some other point of the set $S_I$ 
in its interior, so this cannot occur in the given feasible
decomposition. Therefore, no chain can be associated
with more than two gadget sets. Moreover, a convex chain can
contain at least $k*n^3-O(n)$ points from each of two different
gadget sets without any pivot points $p_x^{(i)}$
in its interior, only if these sets are some 
$B_x^{(i)}$ and $B_x^{(i\pm 1)}$.
(In particular, it is not hard to see that no chain that covers
$\Omega(n)$ points of a set $C_{x,c}$ can cover points
from any other gadget set.)
Since there are $3m+sn=O(n^2)$ chains in total, there
must be at least one chain associated with each
gadget set, and a chain associated with a set $C_{x,c}$
cannot be associated with any other set. This means that
$3m$ of the chains are used to cover the $3m$ chains $C_{x,c}$.
Therefore, the remaining
$2sn$ gadget sets $B_x^{(i)}$ must be covered by the remaining
$sn$ convex chains. None of these chains can cover
more than two of these sets. It follows that the remaining chains form a 
perfect matching on the set of $B_x^{(i)}$, implying that
each chain covers a $B_x^{(i)}$ and a neighboring set $B_x^{(i\pm 1)}$.
Therefore, the gadget sets for each variable are either covered
by pairing along the ``odd'' edges, or by pairing along the ``even''
edges of the associated polygon.

This describes the gadget sets associated with all the convex chains.
Now it is straightforward to verify that a point from one of the triangles
$T_c$ can only be part of one of the chains associated with a
set $C_{x,c}$. This can only happen if this chain does not
intersect a chain along an edge of the polygon for the variable $x$
-- implying that this variable satisfies the clause $c$. 
This completes the proof.
\end{proof}

\begin{theorem}
\label{th:npc.nondisjoint}
It is NP-complete to decide whether for a planar point set $S$
the convex cover number $\kappa_c(S)$ is below some
threshold $k$.
\end{theorem}

\begin{proof}
Our proof uses a reduction of the problem 1-in-3 SAT.
It is inspired by the hardness proof for the {\sc Angular Metric
TSP} given in \cite{ackms-amtsp-97}.

\begin{figure}[hbtp]
   \begin{center}
   \epsfxsize=.35\textwidth
\ \epsfbox{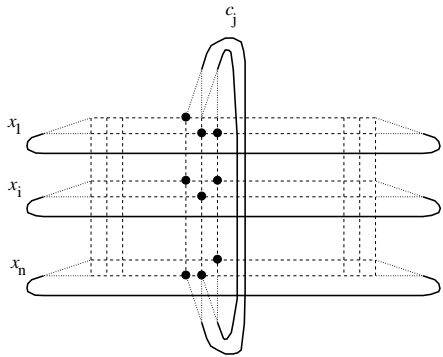}
\caption{A point set $S_I$ for a 1-in-3 SAT instance $I$. Pivot points
are shown for the clause $(x_1\vee \overline{x_i}\vee x_n)$.}
\label{fi:npc.nondisjoint}
   \end{center}
\end{figure}

The construction is as follows:
For a 1-in-3 SAT instance $I$ with $n$ variables
and $m$ clauses, represent each clause
$c_j$ by a triple of vertical columns, each one
associated with a variable that occurs in $c_j$.
Each variable $x_i$ is represented by a pair of 
horizontal rows, the upper one corresponding to ``true'',
the lower one corresponding to ``false''.
This results in a grid pattern
as shown in Figure~\ref{fi:npc.nondisjoint}.
For each pair of a variable $x_i$ and a clause $c_j$, we get
a 2x3 pattern of intersections points. If variable 
$x_i$ appear in clause $c_j$,
we add three ``pivot'' points to this pattern:
If $x_i$ occurs un-negated in $c_j$, a pivot point is 
added at the intersection of $x_i$'s ``true'' row with 
the column of $c_j$ that corresponds
to $x_i$; for both other two columns, a pivot point is added 
at the intersection with the ``false'' row.
If $x_i$ occurs negated in $c_j$, a pivot point is added at the 
intersection of $x_i$'s ``false'' row with the column of $c_j$ 
that corresponds to $x_i$; for both other two columns, a pivot point is 
added at the intersection with the ``true'' row.

Finally, a horizontal ``staple'' gadget that consists
of $\Omega(n^4)$ points is added to the rows for each 
variable, and two nested vertical
staple gadgets are added to the columns for each of the clauses.
These are constructed in a way that a staple forms a
convex chain that can cover all the pivot points in
one row or one column, but not more than that,
and no pivot points from any other clauses or
variables. Thus, for each variable, we can 
collect one of two rows, and for each clause, we can cover
two of three rows.

Now it is possible to show the following:
The 1-in-3 SAT instance $I$ has a satisfying truth assignment,
if and only if the point set $S_I$ can be covered with not more than 
$n+2m$ convex chains. 

It is easy to see that a satisfying truth assignment
implies a small convex cover, by covering each staple gadget
by one chain, and choosing the appropriate
rows and columns of pivot points to be covered by the 
staple gadgets: For each variable, choose
the row corresponding to its truth assignment. For each clause,
choose the two columns for the two variables that do not 
satisfy it. Now it is straightforward to check that all pivot points
are covered.

To see the converse, we can argue in a similar way as in the proof
of Theorem~\ref{th:npc.disjoint} that each staple needs its
own convex chain. Then we are left with a choice of one row
for each variable, and two columns for each clause.
This choice of rows induces a truth assignment to variables;
it is not hard to check that the remaining uncovered pivot points
lie in not more than two columns, if and only if this truth assignment
is valid for the 1-in-3 SAT instance $I$.
\end{proof}

So far, the complexity status of determining the reflexivity of a
point set remains open. However, the apparently close relationship
between convex cover/partition numbers and reflexivity leads us to
believe the following:

\begin{conjecture}
\label{npc.reflex}
It is NP-complete to determine the reflexivity $\rho(S)$
of a point set.
\end{conjecture}

\section{Algorithms}

We have obtained a number of algorithmic results on computing, exactly
or approximately, reflexivity and convex cover/partition numbers.  We
begin with the following theorem, which shows that one can compute
efficiently a constant-factor approximation to the {\em convexivity}
of~$S$:

\begin{theorem}
\label{thm:approx-convexity}
Given a set $S$ of $n$ points in the plane, in $O(n\log n)$ time one
can compute a polygonalization of $S$ having at least 
$\chi(S)/2$
convex vertices, where $\chi(S)=n-\rho(S)$ is the convexivity of~$S$.
\end{theorem}

\begin{proof} 
  The proof of Theorem~\ref{thm:reflex-upper-bnd} is constructive,
  producing a polygonalization of $S$ having at most $\ceil{n_I/2}\leq
  \ceil{n/2}$ reflex vertices, and thus at least $\floor{n/2}$ convex
  vertices (thereby giving a 2-approximation for convexivity).  In
  order to obtain the stated time bound, we must implement the
  algorithm efficiently.  This can be done using a data structure for
  dynamic convex hulls, under pure deletion (see
  Chazelle~\cite{c-clps-85} and Hershberger and
  Suri~\cite{hs-asdch-92}).  At each main step of the algorithm, we
  identify the vertices along the chain $(p_{i+1},q,q_1,\ldots,b)$ by
  making repeated extreme-point queries in the convex hull data
  structure, and then delete the points $p_{i+1},q,q_1,q_2,\ldots$
  from the data structure, and repeat.  The dynamic data structure
  supports deletions and queries in $O(\log n)$ time per operation,
  for an overall time bound of $O(n\log n)$.
\end{proof}

\begin{theorem}
\label{thm:convex-cover-approx}
Given a set $S$ of $n$ points in the plane, the convex cover number,
$\kappa_c(S)$, can be computed approximately, within a factor $O(\log
n)$, in polynomial time.
\end{theorem}

\begin{proof}
  We use a greedy set cover heuristic.  At each stage, we need to
  compute a largest convex subset among the remaining (uncovered)
  points of $S$.  This can be done in polynomial time using the
  dynamic programming methods of~\cite{mrsw-ccppp-95}.
\end{proof}

\begin{theorem}
\label{thm:disjoint-convex-cover-approx}
Given a set $S$ of $n$ points in the plane, the convex
partition number, $\kappa_p(S)$, can be computed approximately, within
a factor $O(\log n)$, in polynomial time.
\end{theorem}

\begin{proof} 
  Let $C^*=\{P_1,\ldots,P_{k^*}\}$ denote an optimal solution,
  consisting of $k^*=\kappa_p(S)$ disjoint convex polygons whose
  vertices are the set $S$.  
  
  By Theorem~2 of~\cite{ers-ccsno-90}, we know that there are $k^*$
  pairwise-disjoint convex polygons $P_1',\ldots,P_{k^*}'$, having a
  total complexity of $O(k^*)$, with $P_i\subseteq P_i'$, for each
  $i$.  Furthermore, the sides of the polygons $P_i'$ can be assumed
  to be segments lying on the $O(n^2)$ lines determined by pairs of
  points of $S$.  Let ${\cal S}$ be the convex polygonal subdivision
  of $C=CH(P_1'\cup\cdots P_{k^*}')$ obtained by decomposing the
  region $C\setminus (P_1'\cup\cdots P_{k^*}')$ into convex polygons
  (e.g., a triangulation suffices).  Then, ${\cal S}$ has total
  complexity $O(k^*)$, its vertices are among the set $V$ of $O(n^4)$
  vertices in the arrangement of $O(n^2)$ lines determined by $S$, and
  its faces are convex polygons.  (Note that some faces of ${\cal S}$
  may be empty of points of $S$.)
  
  We now decompose each face of ${\cal S}$ into a set of vertical
  trapezoids by erecting vertical cuts through the vertices of each
  face, within each face.  (Some of these trapezoids may be triangles,
  which we can consider to be degenerate trapezoids.)  Finally, for
  each such trapezoid $\tau$ we decompose it using vertical cuts into
  $O(\log n^4)=O(\log n)$ {\em canonical trapezoids}, whose
  $x$-projection is one of the $O(n^4)$ canonical $x$-intervals
  determined by a segment tree on $V$.  The resulting canonical
  trapezoidalization, ${\cal T}$, has $O(k^*\log n)$ faces, each of
  which is a canonical trapezoid.  An important property of ${\cal T}$
  is that it has the following binary space partition property: For
  any canonical trapezoid, $\tau$, the subdivision of $\tau$ induced
  by ${\cal T}$ is such that either (a) there exists an edge of ${\cal
    T}$ that cuts $\tau$ in two, extending from its left side to its
  right side, or (b) the vertical cut that splits $\tau$ into two
  canonical subtrapezoids lies entirely on the edge set of ${\cal T}$.
  This property allows us to optimize recursively using dynamic
  programming, in much the same way as was done in
  \cite{as-sagp-98,m-aagsp-93} for problems involving optimal
  separation and surface approximation.
  
  In particular, for any canonical trapezoid, $\tau$, we desire to
  compute the quantity $f(\tau)$, defined to be the minimum number of
  faces in a partitioning of $\tau$ into canonical trapezoids, such
  that within each face of the partitioning, the subset of points of
  $S$ within the face is in convex position.  (The empty set is
  considered trivially to be in convex position.)  Then $f(\tau)$
  obeys a recursion,
  $$f(\tau)=\min\{\min_{\ell}\{f(\tau_\ell^+)+f(\tau_\ell^-)\}, 
  f(\tau_{left})+f(\tau_{right})\},$$
  where the minimization over $\ell$ considers all choices of lines
  $\ell$, intersecting $\tau$ on both of its vertical sides,
  determined by two vertices of $V$; $\tau_\ell^+$ (resp.,
  $\tau_\ell^-$) denotes the portion of $\tau$ lying above (resp.,
  below) the line $\ell$.  We have used $\tau_{left}$ (resp.,
  $\tau_{right}$) to indicate the canonical trapezoid obtained by
  splitting $\tau$ by a vertical line at the $x$-median value (among
  the $x$-coordinates of $V$ that lie in the vertical slab defined
  by~$\tau$).  We leave it to the reader to write the boundary
  conditions of the recursion, which is straightforward.
  
  Our algorithm gives us a minimum-cardinality partition of $S$ into a
  disjoint set, $C'$, of (empty) convex subsets whose $x$-projections
  are canonical intervals.  Since the optimal solution, $C^*$, can be
  converted into at most $k^*\cdot O(\log n)$ such convex sets, we
  know we have obtained an $O(\log n)$-approximate solution to the
  disjoint convex partition problem.
\end{proof}

\begin{corollary}
\label{cor:reflexivity-approx}
Given a set $S$ of $n$ points in the plane, its Steiner reflexivity,
$\rho'(S)$, can be computed approximately, within a factor $O(\log
n)$, in polynomial time.
\end{corollary}

\begin{proof} 
  Let $P^*$ denote an optimal solution, a simple polygon having
  $\rho'(S)$ reflex vertices.  Then, we know that $P^*$ can be
  decomposed into at most $\rho'(S)+1$ convex polygons, each of which
  corresponds to a subset of $S$.  This gives us a partition of $S$
  into at most $\rho'(S)+1$ disjoint convex sets; thus,
  $\kappa_p(S)\leq \rho'(S)+1$.  By
  Theorem~\ref{thm:disjoint-convex-cover-approx}, we can compute a
  set, $C$, of $k\leq O(\log n)\cdot \kappa_p(S)$ disjoint convex
  sets.  We can polygonalize $S$ by ``merging'' these $k$ polygons of
  $C$, using a doubling of a spanning tree on $C$.  The important
  property of the embedding of the spanning tree is that it consists
  of ($k-1$) line segment bridges (with endpoints on the boundaries of
  polygons $C$) that are pairwise non-crossing and do not cross any of
  the polygons $C$. (One way to determine such a tree is to select one
  point interior to each polygon of $C$, compute a minimum spanning
  tree of these $k$ points, and then utilize the portions of the line
  segments that constitute the tree that lie outside of the polygons
  $C$ to be the set of $k-1$ bridging segments.)  A simple
  polygonalization is obtained by traversing the boundary of the union
  of the polygons $C$ and the $k-1$ bridging segments, while slightly
  perturbing the doubled bridge segments.  Since each bridge segment
  is responsible for creating at most 4 new (Steiner) points, this
  results in a polygon, with at most $4(k-1)$ Steiner points, each of
  which may be reflex.  (All other vertices in the polygonalization
  are convex.)  Thus, we obtain a polygonalization with at most
  $4(k-1)\leq O(\log n)\cdot \kappa_p(S)\leq O(\log n)\cdot \rho'(S)$
  reflex vertices.
\end{proof}

For small values of $r$, we have devised particularly efficient
algorithms that check if $\rho(S)\leq r$ and, if so, produce a witness
polygonalization having at most $r$ vertices.  Of course, the case
$r=0$ is trivial, since that is equivalent to testing if $S$ lies in
convex position (which is readily done in $O(n\log n)$ time, which is
worst-case optimal).  It is not surprising that for any
fixed $r$ one can obtain an $n^{O(r)}$ algorithm: enumerate 
over all combinatorially distinct (with respect to $S$) convex subdivisions
of $\textrm{CH}(S)$ into $O(r)$ convex faces and test that the subsets of $S$
within each face are in convex position, and then check all
possible ways to order these $O(r)$ convex chains
to form a circuit that may form a simple polygon.
The factor in front of $r$ in the exponent, however, is not
so trivial to reduce.  In particular, the straightforward method applied
to the case $r=1$ gives $O(n^5)$ time.  With a more careful analysis
of the cases $r=1,2$, we obtain the next two theorems.

\begin{theorem}
\label{thm:small-reflexivity-1}
Given a set $S$ of $n$ points in the plane, in $O(n\log n)$ time one
can determine if $\rho(S)=1$, and, if so, produce a witness
polygonalization.  Further, $\Omega(n\log n)$ is a lower bound on the
time required to determine if $\rho(S)=1$.
\end{theorem}

\begin{figure}
\centerline{\psfig{file=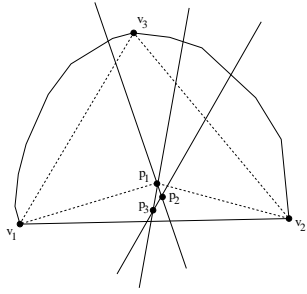,height=1.5in}}
\caption{Proof of the $\Omega(n\log n)$ lower bound
for determining if $\rho(S)=1$.}
\label{fig:lower-bnd-rho1}
\end{figure}

\begin{proof}
  First notice that if a point set $S$ is polygonalized with exactly 1 reflex
  vertex, then (a) the two supporting lines from the endpoints of the lid of
  the pocket to the second convex layer of $S$  are supported
  by the same point $p$ of the layer, which is the reflex vertex of the
  polygonalization, and (b) the vertices of the pocket appear in
  angular order around $p$. The upper bound involves computing the convex
  hull and the second layer of the onion (the entire onion can be computed
  in time $O(n\log n)$), and then performing a careful case
  analysis for how the single pocket must be.

  In fact, if the second convex layer is empty, or has 1 or 2
  points, the solution is trivial. If the second convex layer has three
  points or has all the internal points, then for each edge $e$ of the
  convex hull, trace the supporting lines from its endpoints to the second
  layer. In the cases where the two tangents are supported by the same
  point $p$, check whether or not the angular order of the interior points
  gives a pocket with lid $e$ and only one reflex vertex $p$. The cost of
  this step is $O(n\log n)$: if the layer is a triangle, computing the
  supporting lines can be done in constant time, and the complexity comes
  from sorting the interior points around the vertex $p$ of the
  triangle; if the layer has all the interior points, then the sorted
  order is given, and the complexity comes from the computation of the
  supporting lines (in fact, the supporting lines can be computed in
  overall $O(n)$ time by a ``rotating calipers''-like technique). In all
  the remaining cases, the point set cannot be polygonalized with one reflex
  vertex.
  
  The lower bound follows from convexity testing: determining if a set
  of $n$ points is in convex position.  Given a set $S$ of $n$ points,
  we compute (in $O(n)$ time) one edge, $e=v_1v_2$, of the convex hull
  of $S$.  We then determine (in $O(n)$ time) the point $v_3\in S$
  furthest from the line through $v_1$ and $v_2$; thus, $v_3$ is also
  a vertex of $\textrm{CH}(S)$.
  Next we let $p_1\not\in S$ be a point within $\Delta v_1v_2v_3$ that
  is closer to edge $v_1v_2$ than is any point of $S$.  We also select
  points $p_2$ and $p_3$ within $\Delta p_1v_1v_2$ in such a way that
  $v_3$ lies within the convex cone of apex $p_3$ defined by the rays
  $\vecc{p_3p_1}$ and $\vecc{p_3p_2}$ and that $v_1$ lies within the
  convex cone of apex $p_2$ defined by the rays $\vecc{p_2p_1}$ and
  $\vecc{p_2p_3}$.  (Points $p_1,p_2,p_3$ can be determined in $O(n)$
  time.)  Refer to Figure~\ref{fig:lower-bnd-rho1}.  Then,
  $\vecc{p_3p_1}$ and $\vecc{p_3p_2}$ intersect distinct
  edges of $\textrm{CH}(S)$, as do $\vecc{p_2p_1}$ and $\vecc{p_2p_3}$, while
  $\vecc{p_1p_2}$ and $\vecc{p_1p_3}$ both intersect the edge $v_1v_2$
  of $\textrm{CH}(S)$.  This implies that the only way that a polygonalization
  of $S'=S\cup\{p_1,p_2,p_3\}$ can have only a single reflex vertex is
  if that reflex vertex is $p_1$ and the corresponding pocket has lid
  $v_1v_2$ and vertices $p_1,p_2,p_3$ (with $p_2$ and $p_3$ both being
  convex in the polygonalization).  Thus, the only way to have
  $\rho(S')=1$ is if the points $S$ are in convex position.  We
  conclude that determining if $\rho(S')=1$ is equivalent to solving
  the convex position problem on the input points~$S$.
\end{proof}

\begin{theorem}
\label{thm:small-reflexivity-2}
Given a set $S$ of $n$ points in the plane, in $O(n^3\log n)$ time one
can determine if $\rho(S)=2$, and, if so, produce a witness
polygonalization.
\end{theorem}

\begin{proof}

  We distinguish two cases according to whether or not the two 
reflex vertices belong to the same pocket.

  {\bf Two pockets:}
  Assume that it is possible to polygonalize $S$ with 2 reflex vertices,
  $p$ and $q$, each one in a different pocket.
  Since each pocket only has one reflex vertex, its convex hull
  (including the lid) is a triangle. We distinguish three
  subcases: (i) the two triangles lie entirely on the same side of the
  line $pq$, (ii) the two triangles lie entirely on different sides of
  the line $pq$, and (iii) at least one of the triangles intersects the
  line~$pq$.

  In the first subcase, since all of the interior points lie on one side of
  the line $pq$, the segment $pq$ is an edge of the second convex layer
  of $S$. In addition, the vertices of the pocket containing $p$
  (resp. $q$) appear in angular order about $p$ (resp. $q$). This gives
  a possible algorithm to detect whether such a polygonalization is possible
  for $S$: For each edge $pq$ of the second convex layer of $S$,
  explore all of the points of $S$ in angular order around $p$, starting on
  the side of $pq$ that does not contain any interior point. Once the first
  interior point is found, we have entered the possible pocket of $p$.
  Check the convexity of all (but one) of the interior points found
  before the next external point, which will be the endpoint of the lid
  of the pocket. Proceed symmetrically from $q$ (if the angular order
  around $p$ was checked clockwise, the order around $q$ must be
  checked counter-clockwise). End by making sure that no interior
  points are left unexplored. Since the second convex layer of $S$ has
  $O(n)$ edges, and the checking for each one of them takes linear time,
  this case is checked in overall $O(n^2)$ time.

  In the second subcase, the segment $pq$ may not be an edge of the
  second convex layer, making the previous algorithm impossible
  to apply. On the other hand though, all of the vertices of the
  polygonalization lying on one side of the line $pq$ are angularly sorted
  about $p$, while all of those lying on the other side are angularly sorted
  about $q$. This gives an algorithm to detect whether such a
  polygonalization is possible for $S$. In a first stage we 
  construct a data structure as follows. For each interior point $p$ and
  for each oriented line $\ell$ through $p$, our structure will store
  the following information: (a) the points lying to the left of
  $\ell$, angularly sorted from $p$, (b) a label indicating whether the
  angular order produces a correct polygonalization to the left
  of $\ell$, and (c) in the affirmative, a pointer to the first
  interior point $q$ that $\ell$ will hit when rotated clockwise
  around $p$. This structure can be built in $O(n^2)$ time:
  for each point $p$, it can be initialized at any arbitrary line
  through $p$ in $O(n)$ time, and then all lines through $p$ can be
  explored by rotation around $p$. Every time that a new point
  is found, it is added/eliminated at one end of the ordered list of
  points to the left of $\ell$, checking for convexity of at most one
  vertex (notice that the actualization of the pointer $q$ can be done
  in amortized $O(n)$ time per point $p$).
  The second phase of the algorithm is straightforward. Explore
  all of the pairs $(p,\ell)$ having a satisfactory left
  polygonalization. For each one of them, a pointer indicates its possible
  complementary pair $(q,-\ell)$. Checking whether or not the two
  partial polygonalizations connect properly can be done in constant
  time. Hence, this case can be checked in overall $O(n^2)$ time.

  Finally, in the third subcase, none of the previous good properties
  apply ($pq$ is not necessarily an edge of the second convex
  layer of $S$, and the polygonalization may not be in angular order
  around $p$ or $q$ on one side of the line $pq$), but there is
  at least one side of the line $pq$ where the polygonalization appears
  in angular order both around $p$ and $q$. This gives a possible
  algorithm to detect whether such a polygonalization is possible for
  $S$. Consider all pairs of interior points, $(p,q)$. Angularly check
  around $p$ or $q$ whether all of the points on one side of the line
  $pq$ form a convex chain. If so, keep turning around $p$ and
  around $q$ separately and in opposite senses on the other
  halfplane, until the two possible pockets are found. The
  remaining chain between $p$ and $q$ can be checked from any of the
  two points. Since we perform linear time work for each pair of
  interior points, this case is checked in overall $O(n^3)$ time.

  {\bf One pocket:}
  Assume that it is possible to polygonalize $S$ with 2 reflex vertices,
  $p$ and $q$, both belonging to the same pocket, with lid $ab$.
  The pocket is formed by three convex chains: $ap$, $pq$ and
  $qb$. We will distinguish two subcases, depending on whether or not the
  chains $ap$ and $qb$ are separable by a line through
  the point $p$. In the first subcase, the vertices of the chain $ap$
  appear in angular order around $p$ before finding any point of the
  remaining chains. In the second subcase, since the chains $ap$ and $qb$
  are convex, they are linearly separable by a line defined by one point
  $l\in ap$ and one point $r\in qb$. Such a line must intersect the lid
  of the pocket.
  In addition, we will distinguish the subcase in which the convex hull of
  the pocket, including the lid endpoints $a$ and $b$ is a
  quadrilateral from the case in which it is a triangle. In the fist subcase,
  the segment $pq$ is an edge of the second convex layer of $S$, while
  in the second subcase it is not. These observations give an
  algorithm to detect whether such a polygonalization is possible for~$S$.

  The subcase {\em separable-quadrilateral} can be detected in $O(n^2)$ time.
  For each edge $pq$ of the second convex layer of $S$, compute the
  lid $ab$ of the possible pocket, if it exists, by intersecting the
  convex hull of $S$ with the prolongations of the edges of the second
  layer incident in $p$ and $q$. From $p$, explore in angular order the
  interior points until the first left turn is reached. Then
  check whether the remaining interior points behave properly when
  explored in angular order from~$q$.

  The subcase {\em separable-triangle} can be detected in $O(n^2\log n)$ time.
  For each edge $ab$ of the convex hull, find the reflex vertex $p$
  of the possible pocket, if it exists, by computing the supporting
  lines from $a$ and $b$ to the second convex layer of $S$.
  From $p$, explore in angular order the interior points until the
  first left turn is reached. Then, check whether the remaining
  interior points, together with $p$ and $b$, form a set $S'$ such
  that $\rho(S')=1$. In this case we do not have a candidate $q$ to
  help us, and the complexity of this procedure comes from
  Theorem~\ref{thm:small-reflexivity-1}.
  
  The subcase {\em not-separable-quadrilateral} can be detected in
  $O(n^3)$ time.  For each pair $(l,r)$ of points of $S$, intersect
  the line $lr$ with the convex hull of $S$ to compute the two
  possible lids. Let $ab$ be a possible lid for $lr$; we will call $s$
  the opposite intersection point of $lr$ and the convex hull of $S$.
  Compute the supporting lines from $a$ and $b$ to the second convex
  layer to obtain the candidate points $p$ and $q$ associated with the
  lid. Explore all of the points to the left of $lr$, together with
  $s$ and $l$ (resp. $r$), in angular order around $p$. Analogously,
  do this for the points to the right, around $q$. If a suitable
  polygonalization is possible in each halfplane, check the connection
  between them. For each of the $O(n^2)$ pairs of points $lr$, we have
  performed linear time work.
  
  The subcase {\em not-separable-triangle} can be detected in
  $O(n^3\log n)$ time.  The only difference from the previous subcase
  is that, to the right of the line $lr$ we do not have a point $q$ to
  be used to perform the checking in angular order. But we can check
  whether the interior points to the right of $lr$, together with $b$
  and $s$, form a set $S'$ such that $\rho(S')=1$. The complexity of
  this procedure ($O(n\log n)$) comes from
  Theorem~\ref{thm:small-reflexivity-1}.
\end{proof}

\section{Inflectionality of Point Sets}
\label{sec:inflection}

Consider a clockwise traversal of a polygonalization, $P$, of $S$.
Then, convex (resp., reflex) vertices of $P$ correspond to {\em right}
(resp., {\em left}) turns.  In computing the reflexivity of $S$ we
desire a polygonalization that minimizes the number of left turns.  In
this section we consider the related problem in which we want to
minimize the number of {\em changes} between left-turning and
right-turning during a traversal that starts (and ends) at a point
interior to an edge of $P$.  We define the minimum number of such
transitions between left and right turns to be the {\em
  inflectionality}, $\phi(S)$, of $S$, where the minimum is taken over
all polygonalizations of $S$. 
(An alternative definition is based on
defining an {\em inflection edge} of $P$ to be an edge connecting a
reflex vertex and a convex vertex; the inflectionality is the minimum
number of inflection edges in any polygonalization of $S$.)
Clearly, $\phi(S)$ must be an even integer; it is zero if and only if $S$
is in convex position.  Somewhat surprisingly, it turns out that
$\phi(S)$ can only take on the values 0 or~2:

\begin{theorem}
\label{th:inflex}
For any finite set $S$ of $n$ points in the plane, $\phi(S)\in
\{0,2\}$, with $\phi(S)=0$ precisely when $S$ is in convex position.
In $O(n\log n)$ time, one can determine $\phi(S)$ as well as a
polygonalization that achieves inflectionality~$\phi(S)$.
\end{theorem}

\begin{proof}
  If $S$ is in convex position, then trivially $\phi(S)=0$.  Thus,
  assume that $S$ is not in convex position.  Then $\phi(S)\neq 0$, so
  $\phi(S)\geq 2$.  We claim that $\phi(S)=2$.  For simplicity, we
  assume that $S$ is in general position.
  
  Consider the $\ell$ nested convex polygons, $C_1,C_2,\ldots,C_\ell$,
  whose boundaries constitute the $\ell$ layers (the ``onion'') of the
  set $S$; these can be computed in time $O(n\log
  n)$~\cite{c-clps-85}.
  
  We construct a ``spiral'' polygonalization of $S$ based on taking
  one edge, $ab$, of $C_1$, and replacing it with a pair of
  right-turning chains from $a$ to $p\in S\cap C_\ell$ and from $b$ to
  $p$.  The two chains exactly cover the points of $S$ on layers
  $C_2,\ldots,C_\ell$.  A constructive proof that such a
  polygonalization exists is based on the following claim:

\begin{claim}
  For any $1\leq m<\ell$ and any pair, $a,b\in S$, of
  vertices of $C_m$, there exist two purely right-turning chains,
  $\gamma_a=(a,u_1,u_2,\ldots,u_i,p)$ and
  $\gamma_b=(b,v_1,v_2,\ldots,v_j,p)$, such that the points of $S$
  interior to $C_m$ are precisely the set $\{u_1,u_2,\ldots,u_i, p,
  v_i,v_2,\ldots,v_j\}$.
\end{claim}

\begin{proof}[Proof of Claim]
We prove the claim by induction on $m$.  If $m=\ell-1$,
the claim follows easily, by a case analysis
as illustrated in Figure~\ref{fig:easy-case}.

\begin{figure}[hbtp]
   \begin{center}
   \epsfxsize=.85\textwidth
\ \epsfbox{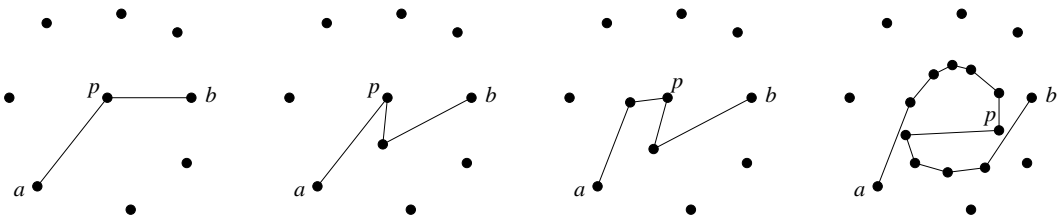}
\caption{Simple case in the inductive proof: $m=\ell-1$.  There are
four subcases, left to right: (i)
$C_\ell$ is a single point; (ii) $C_\ell$ is a line segment
determined by two points of $S$; (iii) $C_\ell$ is a triangle
determined by three points of $S$; or (iv) $C_\ell$ is a convex polygon
whose boundary contains four or more points of $S$.}
\label{fig:easy-case}
   \end{center}
\end{figure}

Assume that the claim holds for $m\geq k+1$ and consider the case
$m=k$.  If $C_{k+1}$ is either a single vertex or a line segment
(which can only happen if $k+1=\ell$), the claim trivially follows;
thus, we assume that $C_{k+1}$ has at least three vertices.  We
let $u_1$ be the vertex of $C_{k+1}$ that is a {\em left tangent}
vertex with respect to $a$ (meaning that $C_{k+1}$ lies in the closed
halfplane to the right of the oriented line $au_1$); we let $v$ be the
left tangent vertex of $C_{k+1}$ with respect to $b$.  
Refer to Figure~\ref{fig:inflectionality-proof2}.
If $v=u_1$, we
define $v_1$ to be the vertex of $C_{k+1}$ that is the
counter-clockwise neighbor of $u_1$; otherwise, we let $v_1=v$.  Let
$a'$ be the counter-clockwise neighbor of $v_1$.  Let $b'$ be the
counter-clockwise neighbor of $u_1$.  (Thus, $b'$ may be the same
point as $v_1$.)  By the induction hypothesis, we know that there
exist right-turning chains, $\gamma_{a'}$ and $\gamma_{b'}$, starting
from the points $a'$ and $b'$, spiraling inwards to a point $p$
interior to $C_{k+1}$.  Then we construct $\gamma_a$ to be the chain
from $a$ to $u_1$, around the boundary of $C_{k+1}$ clockwise to $a'$,
and then along the chain $\gamma_{a'}$.  Similarly, we construct
$\gamma_b$ to be the chain from $b$ to $v_1$, around the boundary of
$C_{k+1}$ clockwise to $b'$, and then along the chain $\gamma_{b'}$.
\end{proof}

\begin{figure}[hbtp]
\centerline{\hfill\psfig{file=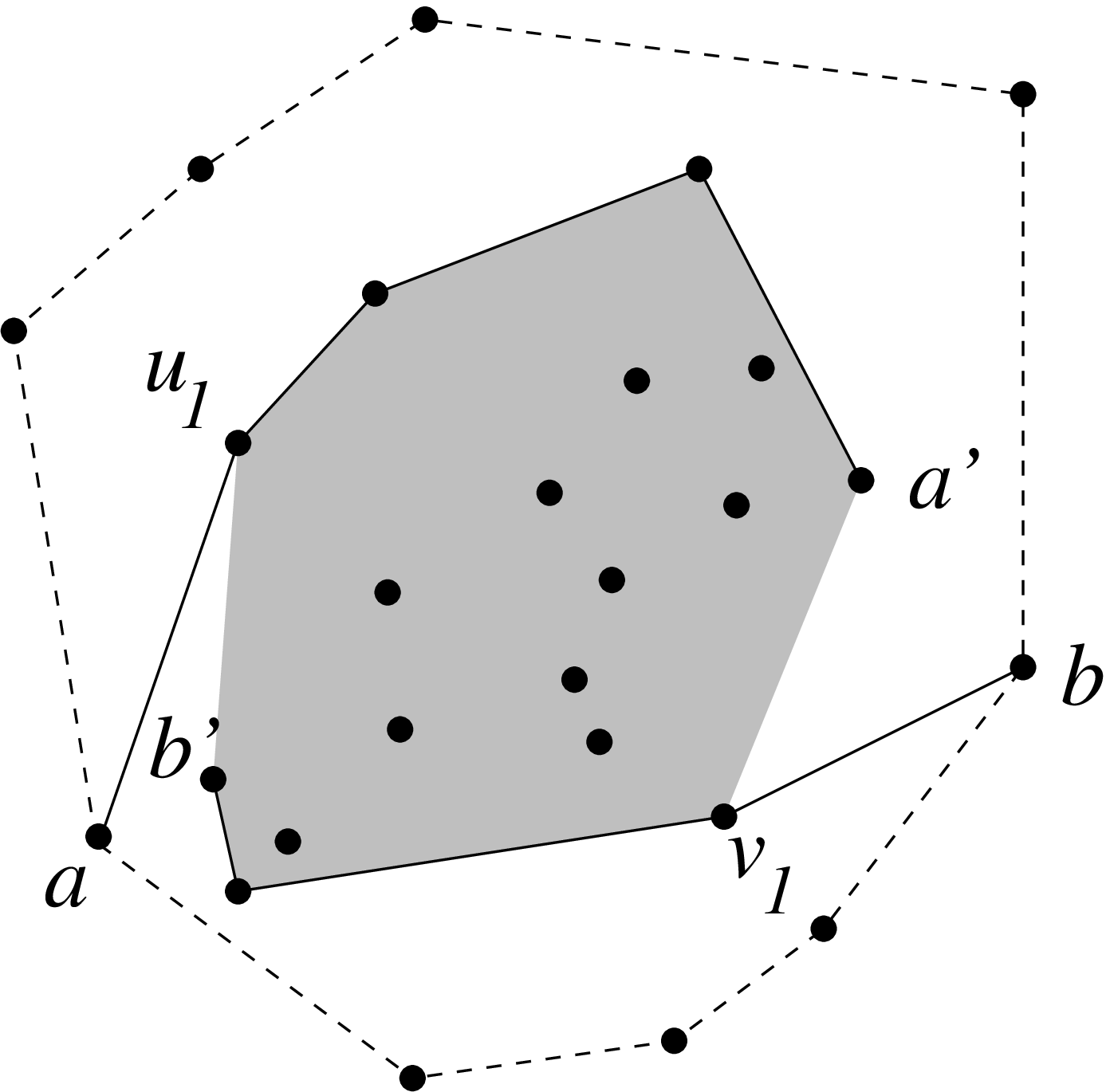, width=0.25\textwidth}\hfill 
\psfig{file=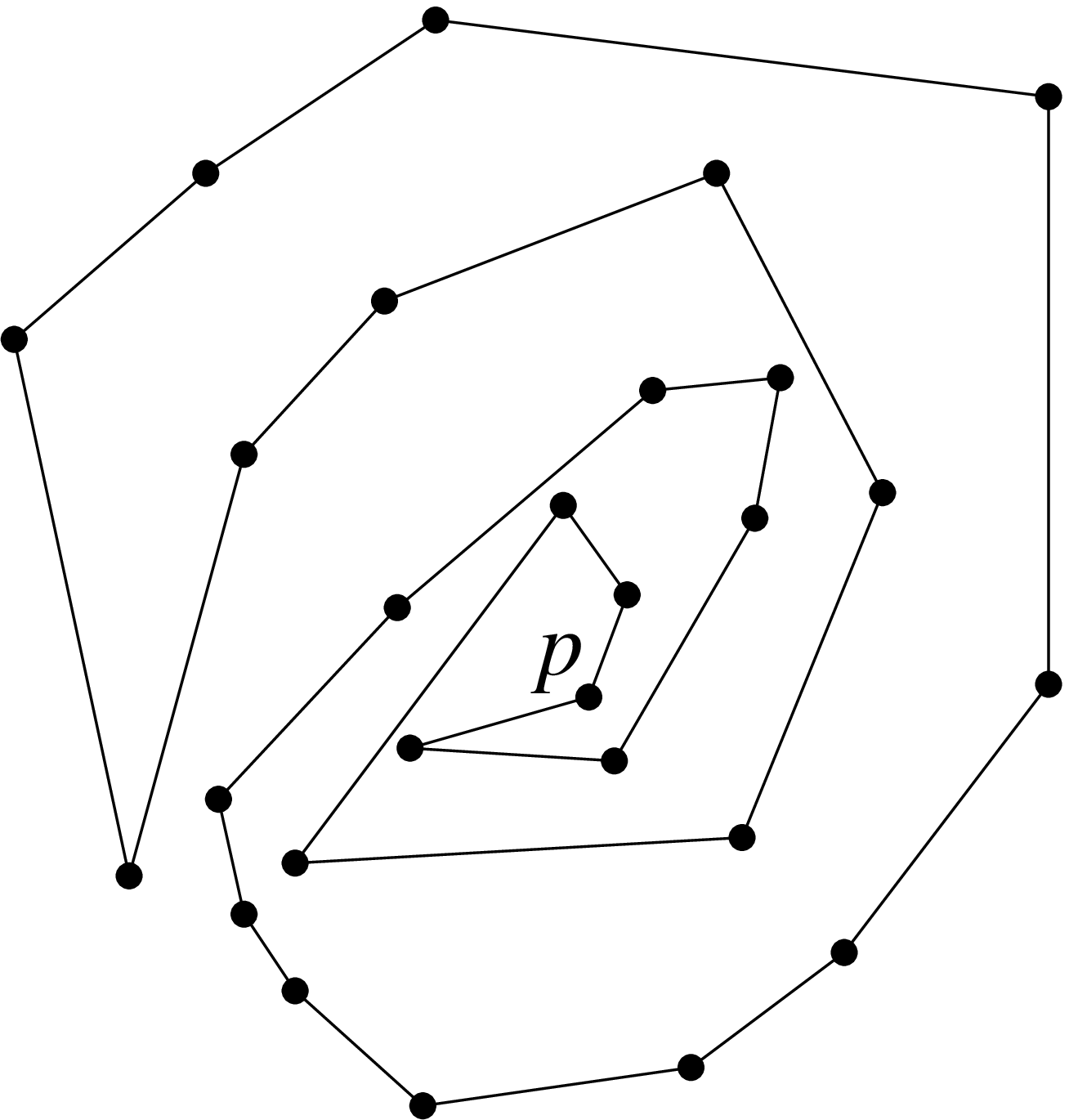, width=0.25\textwidth}\hfill}
\caption{Left: Constructing the spiraling chains $\gamma_a$ and $\gamma_b$.
Right: An example of the resulting spiral polygonalization.}
\label{fig:inflectionality-proof2}
\end{figure}

The proof of the above claim is constructive; the
required chains are readily obtained in $O(n\log n)$ time,
given the convex layers.  This concludes the proof of the theorem.
\end{proof}

\section{Conclusion and Future Work}

We have introduced a new class of combinatorial and algorithmic
problems related to simple polygonalizations of a planar point set.
We have given lower and upper combinatorial bounds, settled
the complexity status of some problem variants, and given
some efficient algorithms, both exact algorithms and
approximation algorithms.

There are a number of interesting open problems that our work
suggests.  First, there are the four specific conjectures mentioned
throughout the paper; these represent to us the most outstanding open
questions raised by our work.  In addition, we mention three other areas
of future study:
\begin{enumerate}
\item Instead of minimizing the number of reflex vertices, can we
  compute a polygonalization of $S$ that minimizes the sum of the {\em
    turn angles} at reflex vertices?  (The turn angle at a reflex
  vertex having interior angle $\theta>\pi$ is defined to be
  $\theta-\pi$.)  This question was posed to us by Ulrik Brandes.  It
  may capture a notion of goodness of a polygonalization that is
  useful for curve reconstruction.  The problem differs from the
  angular metric TSP (\cite{ackms-amtsp-97}) in that the only turn
  angles contributing to the objective function are those of reflex
  vertices.
\item What can be said about the generalization of the reflexivity
  problem to polyhedral surfaces in three dimensions?  This may be of
  particular interest in the context of surface reconstruction.
\item There are a number of natural measures of ``near convexity'' for
  point sets.  It would be interesting to do a systematic study of how
  the various measures compare.
\end{enumerate}

\subsection*{Acknowledgments}

We thank Adrian Dumitrescu for valuable input on this work, including
a software tool for calculating reflexivity of point sets.  We thank
Oswin Aichholzer for applying his software to search all
combinatorially distinct small point sets.  This collaborative
research between the Universitat Polit\`ecnica de Catalunya and Stony
Brook University was made possible by a grant from the Joint
Commission USA-Spain for Scientific and Technological Cooperation
Project 98191.  E. Arkin acknowledges additional support from the
National Science Foundation (CCR-9732221, CCR-0098172).  S.\ Fekete
acknowledges travel support by the Hermann-Minkowski-Minerva Center
for Geometry at Tel Aviv University.  F.\ Hurtado, M.\ Noy, and V.\ 
Sacrist\'an acknowledge support from CUR Gen. Cat. 1999SGR00356, and
Proyecto DGES-MEC PB98-0933.  J.~Mitchell acknowledges support from
NSF (CCR-9732221, CCR-0098172) and NASA Ames Research Center
(NAG2-1325).

\nocite{ackms-amtsp-97,u-ppsdc-99,u-opcp-96,u-ppscp-97}
\nocite{es-sepeg-60,es-cpg-35,deo-secp-90,gnt-lbncf-95}

\bibliographystyle{abbrv}

\begin{thebibliography}{10}

\bibitem{a-rsoas-92}
P.~K. Agarwal.
\newblock Ray shooting and other applications of spanning trees with low
  stabbing number.
\newblock {\em SIAM J. Comput.}, 21:540--570, 1992.

\bibitem{afg-pdecm-02}
P.~K. Agarwal, E. Flato, and D. Halperin.
Polygon decomposition for efficient
construction of Minkowski sums.
\newblock {\em Comput. Geom. Theory Appl.}, 21:39--61, 2002.

\bibitem{as-sagp-98}
P.~K. Agarwal and S.~Suri.
\newblock Surface approximation and geometric partitions.
\newblock {\em SIAM J. Comput.}, 27:1016--1035, 1998.

\bibitem{ackms-amtsp-97}
A.~Aggarwal, D.~Coppersmith, S.~Khanna, R.~Motwani, and B.~Schieber.
\newblock The angular-metric traveling salesman problem.
\newblock In {\em Proceedings of the Eighth Annual ACM-SIAM Symposium on
  Discrete Algorithms}, pages 221--229, Jan. 1997.

\bibitem{aak-eotsp-01}
O.~Aichholzer, F.~Aurenhammer, and H. Krasser.
\newblock Enumerating order types for small point sets with applications.
\newblock In {\em Proc. 17th Annu. ACM Sympos. Comput. Geom.}, 2001, pp.~11--18.

\bibitem{ak-psotd-01}
O. Aichholzer and H. Krasser.
\newblock The point set order type data base: a collection of applications
and results. 
\newblock In {\em Proc. 13th Canad. Conf. Comput.
Geom.}, Waterloo, Canada, 2001, pp.~17--20.

\bibitem{AK}
O.~Aichholzer and H.~Krasser.
\newblock Personal communication, 2001.

\bibitem{abe-cbscc-98}
N.~Amenta, M.~Bern, and D.~Eppstein.
\newblock The crust and the $\beta$-skeleton: Combinatorial curve
  reconstruction.
\newblock {\em Graphical Models and Image Processing}, 60:125--135, 1998.

\bibitem{ah-hgrp-96}
T.~Auer and M.~Held.
\newblock Heuristics for the generation of random polygons.
\newblock In {\em Proc. 8th Canad. Conf. Comput. Geom.}, pages 38--43, 1996.

\bibitem{c-cgc-79}
B.~Chazelle.
\newblock {\em Computational geometry and convexity}.
\newblock Ph.{D}. thesis, Dept. Comput. Sci., Yale Univ., New Haven, CT, 1979.
\newblock Carnegie-Mellon Univ. Report CS-80-150.

\bibitem{c-clps-85}
B.~Chazelle.
\newblock On the convex layers of a planar set.
\newblock {\em IEEE Trans. Inform. Theory}, IT-31(4):509--517, July 1985.

\bibitem{dk-spacr-99}
T.~K. Dey and P.~Kumar.
\newblock A simple provable algorithm for curve reconstruction.
\newblock In {\em Proc. 10th ACM-SIAM Sympos. Discrete Algorithms}, pages
  893--894, Jan. 1999.

\bibitem{dmr-crcdgr-99}
T.~K. Dey, K.~Mehlhorn, and E.~A. Ramos.
\newblock Curve reconstruction: {C}onnecting dots with good reason.
\newblock In {\em Proc. 15th Annu. ACM Sympos. Comput. Geom.}, pages 197--206,
  1999.

\bibitem{deo-secp-90}
D.~P. Dobkin, H.~Edelsbrunner, and M.~H. Overmars.
\newblock Searching for empty convex polygons.
\newblock {\em Algorithmica}, 5:561--571, 1990.

\bibitem{ers-ccsno-90}
H.~Edelsbrunner, A.~D. Robison, and X.~Shen.
\newblock Covering convex sets with non-overlapping polygons.
\newblock {\em Discrete Math.}, 81:153--164, 1990.

\bibitem{es-cpg-35}
P.~Erd{\H o}s and G.~Szekeres.
\newblock A combinatorial problem in geometry.
\newblock {\em Compositio Math.}, 2:463--470, 1935.

\bibitem{es-sepeg-60}
P.~Erd{\H o}s and G.~Szekeres.
\newblock On some extremeum problem in geometry.
\newblock {\em Ann. Univ. Sci. Budapest}, 3-4:53--62, 1960.

\bibitem{e-webpage}
J. Erickson.
Generating random simple polygons.\\
{\tt http://compgeom.cs.uiuc.edu/\~{}jeffe/open/randompoly.html}

\bibitem{fw-artp-97}
S.~P. Fekete and G.~J. Woeginger.
\newblock Angle-restricted tours in the plane.
\newblock {\em Comput. Geom. Theory Appl.}, 8(4):195--218, 1997.

\bibitem{gnt-lbncf-95}
A.~Garc{\'\i}a, M.~Noy, and J.~Tejel.
\newblock Lower bounds for the number of crossing-free subgraphs of {$K_n$}.
\newblock In {\em Proc. 7th Canad. Conf. Comput. Geom.}, pages 97--102, 1995.

\bibitem{hs-asdch-92}
J.~Hershberger and S.~Suri.
\newblock Applications of a semi-dynamic convex hull algorithm.
\newblock {\em BIT}, 32:249--267, 1992.

\bibitem{hs-parss-95}
J.~Hershberger and S.~Suri.
\newblock A pedestrian approach to ray shooting: {Shoot} a ray, take a walk.
\newblock {\em J. Algorithms}, 18:403--431, 1995.

\bibitem{hm-ftprs-85}
S.~Hertel and K.~Mehlhorn.
\newblock Fast triangulation of the plane with respect to simple polygons.
\newblock {\em Inform. Control}, 64:52--76, 1985.

\bibitem{h-snec7g-83}
J.~Horton.
\newblock Sets with no empty convex 7-gons.
\newblock {\em Canad. Math. Bull.}, 26:482--484, 1983.

\bibitem{HRU}
K.~Hosono, D.~Rappaport, and M.~Urabe.
\newblock On convex decompositions of points.
\newblock In {\em Proc. Japanese Conf. on Discr. Comp. Geom. (2000)}, volume
  2098 of {\em Lecture Notes Comput. Sci.}, pages 149--155. Springer-Verlag,
  2001.

\bibitem{hu-ndcqp-01}
K.~Hosono and M.~Urabe.
\newblock On the number of disjoint convex quadrilaterals for a plannar point
  set.
\newblock {\em Comp. Geom. Theory Appl.}, 20:97--104, 2001.

\bibitem{hn-tvgrv-96}
F.~Hurtado and M.~Noy.
\newblock Triangulations, visibility graph and reflex vertices of a simple
  polygon.
\newblock {\em Comput. Geom. Theory Appl.}, 6:355--369, 1996.

\bibitem{k-pd-00}
J.~M. Keil.
\newblock Polygon decomposition.
\newblock In J.-R. Sack and J.~Urrutia, editors, {\em Handbook of Computational
  Geometry}, pages 491--518. Elsevier Science Publishers B.V. North-Holland,
  Amsterdam, 2000.

\bibitem{l-pftu-82}
D.~Lichtenstein.
\newblock Planar formulae and their uses.
\newblock {\em SIAM J. Comput.}, 11(2):329--343, 1982.

\bibitem{m-aagsp-93}
J.~S.~B. Mitchell.
\newblock Approximation algorithms for geometric separation problems.
\newblock Technical report, Department of Applied Mathematics, SUNY Stony
  Brook, NY, July 1993.

\bibitem{mrsw-ccppp-95}
J.~S.~B. Mitchell, G.~Rote, G.~Sundaram, and G.~Woeginger.
\newblock Counting convex polygons in planar point sets.
\newblock {\em Inform. Process. Lett.}, 56:191--194, 1995.

\bibitem{o-cgc-98}
J.~O'Rourke.
\newblock {\em Computational Geometry in {C}}.
\newblock Cambridge University Press, 2nd edition, 1998.

\bibitem{dcg19}
J.~Pach, editor.
\newblock {\em Special Issue Dedicated to Paul {Erd\"{o}s}}, volume~19 of {\em
  Discrete Comput. Geom.}
\newblock 1998.

\bibitem{RU}
E.~Rivera-Campo and J.~Urrutia.
\newblock Personal communication, 2001.

\bibitem{u-opcp-96}
M.~Urabe.
\newblock On a partition into convex polygons.
\newblock {\em Discrete Appl. Math.}, 64:179--191, 1996.

\bibitem{u-ppscp-97}
M.~Urabe.
\newblock On a partition of point sets into convex polygons.
\newblock In {\em Proc. 9th Canad. Conf. Comput. Geom.}, pages 21--24, 1997.

\bibitem{u-ppsdc-99}
M.~Urabe.
\newblock Partitioning point sets into disjoint convex polytopes.
\newblock {\em Comput. Geom. Theory Appl.}, 13:173--178, 1999.

\bibitem{w-ggssn-93}
E.~Welzl.
\newblock Geometric graphs with small stabbing numbers: {Combinatorics} and
  applications.
\newblock In {\em Proc. 9th Internat. Conf. Fund. Comput. Theory}, Lecture
  Notes Comput. Sci., Springer-Verlag, 1993.

\bibitem{zssm-grpgv-96}
C.~Zhu, G.~Sundaram, J.~Snoeyink, and J.~S.~B. Mitchell.
\newblock Generating random polygons with given vertices.
\newblock {\em Comput. Geom. Theory Appl.}, 6:277--290, 1996.

\end{thebibliography}

\end{document}